\documentclass[12pt,preprint]{aastex}

\renewcommand\>{{\rangle}}
\newcommand\bld[1]{\mbox{\boldmath $#1$}}
\newcommand\<{{\langle}}

\newcommand{\pdv}[2]{\frac{\partial#1}{\partial#2}}
\newcommand{\bnabla}{\bld{\nabla}}
\newcommand{\bB}{\bld{B}}

\newcommand{\bx}{\bld{x}}
\newcommand{\by}{\bld{y}}

\renewcommand{\bv}{\bld {v}}

\newcommand{\bO}{\bld{\Omega}}
\newcommand{\uv}[1]{\hat{\bld{#1}}}
\newcommand{\vorb}{\bv_{orb}}

\newcommand{\athenap}{{\tt ATHENA}}
\newcommand{\zeus}{{\tt ZEUS }}
\newcommand{\zeusp}{{\tt ZEUS}}

\newcommand{\del}{\partial}
\newcommand{\pr}{{\rm\, Pr}}

\newcommand\msun{{\rm\,M_\odot}}

\newcommand{\be}{\begin{equation}}
\newcommand{\ee}{\end{equation}}
\newcommand{\bea}{\begin{eqnarray}}
\newcommand{\eea}{\end{eqnarray}}

\newcommand\abe{\texttt{\bfseries abe}}
\newcommand\queenbee{\texttt{\bfseries queenbee}}
\newcommand\ranger{\texttt{\bfseries ranger}}
\newcommand\kraken{\texttt{\bfseries kraken}}
\shortauthors{Guan and Gammie}
\shorttitle{}
\begin{document}

\title{Radially Extended, Stratified, Local Models of Isothermal Disks}

\author{Xiaoyue Guan\altaffilmark{1} and Charles F. Gammie\altaffilmark{2}}

\altaffiltext{1}{Astronomy Department, University of Virginia}
\altaffiltext{2}{Physics Department, University of Illinois}

\begin{abstract}

We consider local, stratified, numerical models of isothermal accretion
disks.  The novel feature of our treatment is that radial extent $L_x$
and azimuthal extent $L_y$ satisfy $H \ll L_x, L_y \ll R$, where $H$ is
the scale height and $R$ is the local radius.  This enables us to probe {\em mesoscale} structure in stratified thin disks.
We evolve the model at several resolutions, sizes, and initial magnetic
field strengths.  Consistent with earlier work, we find that the
saturated, turbulent state consists of a weakly magnetized disk midplane
coupled to a strongly magnetized corona, with a transition at $|z| \sim
2 H$.  The saturated $\alpha \simeq 0.01 - 0.02$.  A two-point
correlation function analysis reveals that the central $4 H$ of the disk
is dominated by small scale turbulence that is statistically similar to
unstratified disk models, while the coronal magnetic fields are
correlated on scales $\sim 10 H$.  Nevertheless angular momentum
transport through the corona is small.  A study of magnetic field loops
in the corona reveals few open field lines and predominantly toroidal
loops with a characteristic distance between footpoints that is $\sim
H$.  Finally we find quasi-periodic oscillations with characteristic
timescale $\sim 30 \Omega^{-1}$ in the magnetic field energy density.
These oscillations are correlated with oscillations in the mean
azimuthal field; we present a phenomenological, alpha-dynamo model that
captures most aspects of the oscillations.

\end{abstract}


\section{Introduction}

The physics of angular momentum transport is at the core of accretion
disk studies. Classical viscous thin disk theories (\cite{ss, lbp,
nt73}) assume the existence of a local turbulent viscous stress, thus
provide a simple local parameterization, i.e., ``anomalous viscosity"
$\alpha$, for disk momentum transport and dissipation. Since the early
90's, the magnetorotational instability (MRI, \cite{bh91}) has been
regarded as the best candidate to drive accretion disk turbulence,
although gravitational torque or magnetic winds of a \cite{bp82} type
can also enhance angular momentum transport.

Classical thin disk theories are vertically integrated and azimuthally
averaged, therefore essentially one dimensional. Currently, disk
vertical structure can only be obtained from numerical simulations where
turbulence is established from first-principle instabilities such as the
MRI. Global disk simulations are just starting to investigate thin disks
(\cite{rf08, shafee08, nkh09}), but they are computationally expensive
and not yet able to fully resolve turbulent structures in the disk.
Shearing box simulations, on the other hand, can concentrate resolution
on disk dynamics at scales of order the disk scale height $H \equiv
c_s/\Omega$, therefore are more suitable to study accretion flows in
detail. Past studies of shearing box simulations with vertical
gravity (e.g., \cite{bnst95, shgb96, ms00, hks06, bhk07, si09}) have
revealed a rich set of structures and dynamics in stratified disks.
However, all these stratified shearing box simulations were done with a
box of limited radial extent $L_x \sim H$, therefore they were not able
to explore any structure on scales larger than $H$. Recently,
\cite{dsp10} have studied stratified shearing box of radial extent $L_x =
4H$, and \cite{jky09} have adopted models of box size up to $L_x \sim 10H$ in
their zonal flow studies. However both these studies are limited to
the small veritical extent ($\sim \pm 2H$) and physically unrealistic periodic vertical boundary conditions. In this paper we
study the dynamics and structure in isothermal stratified disks using
large shearing box with domain sizes $L_x \geq 10 H$ in all directions. 

We still do not know whether a magnetized turbulent disk is well modeled
as a steady-state, locally dissipated disk model. It is possible, for
example, that structures (gas and/or fields) develop at a scale large
compared to $H$, and that these structures could be associated with
nonlocal energy or angular momentum transport. Large scale structures
might also develop in the magnetic field in the form of dynamo.  The
disk might also be secularly unstable (see the overview by \cite{pir78}),
that could cause the disk to break up into rings.  It is well known
that a Navier-Stokes viscosity model for disk turbulence leads, for some
opacity regimes, to both viscous \citep{le74} and thermal \cite{pir78}
instability, although it is now believed that thermal instability can be
removed by delays imposed through finite relaxation time effects in
MRI-driven turbulence (\cite{hkb09}).

>From an observational point of view, the level of fluctuations
(inhomogeneity) at different locations in disks and how these different
locations communicate with each other have important consequences for
disk spectra modeling \citep{dbht05, bdhks06}. In these models
observational diagnostics require integrating over the disk surface, so
radially extended structure in the disk model may change the disk
spectrum.  Our disk model is isothermal (we do not solve an energy
equation) and is therefore not capable of investigating dissipation and
radiation.  It is possible that larger fluctuations would appear in
physically richer models where thermodynamics and radiative effects are
taken into account (e.g., \cite{tsks03, turner04}).  It would then be
interesting in the future for spectral modelers to consider disk models
with larger radial domains.  

A shearing box larger than $H$ is also essential to catch the field
structure and dynamics in the accretion disk magnetic coronae (ADC;
\cite{tp96}; also see a discussion in \cite{uzgoo}), where the field has
a characteristic curvature $l \sim v_a/\Omega \geq H$, and $v_a$ is the
characteristic Alfv\'en speed in the region. 

Recently, it has also been pointed out that a large box size may be
important to study the saturation properties of the MRI-driven
turbulence, either on the ground of resolving parasitic modes
\citep{pg09}, or in a phenomenological model of an MRI driven dynamo
\citep{v09}. Saturation mechanisms in stratified disk may be
fundamentally different from those in unstratified disks.  Recent
numerical experiments on unstratified disks suggest that: (a) with a
zero-net flux, the saturation is dependent on the microscopic Prandtl
number $\pr_M$ in the disk, at least at the low Reynolds number
\citep{fp07,ll07, sh09}; (b) with a net (toroidal or vertical) flux, the
saturation increases with resolution \citep{hgb95, ggsj09}. Stratified
disk models, which are closer to real disks, may well maintain a net
(most likely, toroidal; see a discussion in \cite{gg09}) field in the
disk region because of the magnetic buoyancy induced by stratification.
Therefore we expect a saturation in stratified disk models to differ
from unstratified models. 

It is worth enumerating the assumptions we adopt in this work: (1) we
use an isothermal equation of state (EOS) in our models; (2) the
vertical support comes from the gas and magnetic pressure rather than
the radiation pressure; (3) there is no explicit viscosity or
resistivity; (4) our initial conditions consist of a uniform toroidal
field in a region near the disk midplane; (5) we use outflow boundary
conditions for the vertical boundaries.

The paper is organized as follows. In \S 2 we give a description of the
local model and summarize our numerical algorithm. In \S 3 we present a
fiducial model and analyze its structure in the saturated state.  In \S
4 we describe how this structure depends on model parameters.  In \S 5
we give a report on quasiperiodic oscillations (``butterfly diagrams'')
and present a phenomenological model to describe them that is based on a
mean-field dynamo model;  \S 5 contains a summary of our results.

\section{Local Model and Numerical Methods}

The local model for disks can be obtained by expanding the equations of
motion around a circular-orbiting coordinate origin, with $(r,\phi,z) =
(r_o, \Omega_o t + \phi_o, 0)$ in cylindrical coordinates, assuming that
the peculiar velocities are comparable to the sound speed and that the
sound speed is small compared to the orbital velocity.  The local
Cartesian coordinates are then obtained from cylindrical coordinates via
$(x,y,z) = (r - r_o, r_o [\phi - \Omega_o t - \phi_o], z)$.  In this
work we assume the disk sits in a Keplerian ($1/r$) potential. We
also use an isothermal ($p = c_s^2 \rho$, where $c_s$ is constant) EOS.

For an ideal MHD disk, the equation of motion in the local model is
\be\label{BE2}
\pdv{\bv}{t} + \bv\cdot \bnabla \bv + c_s^2\frac{\bnabla\rho}{\rho} + \frac{\bnabla B^2}{8\pi \rho}
- \frac{(\bB\cdot \bnabla)\bB}{4\pi \rho} + 2 \bO \times \bv -
3\Omega^2 x \, \uv{x}  + \Omega^2 z \,\uv{z} = 0.
\ee
>From left to right, the last three terms in Eqn (\ref{BE2}) represent
the Coriolis force, tidal forces and vertical gravitational acceleration
in the local frame respectively. The orbital velocity in the local model
is
\be
\vorb = -{3\over{2}}\Omega x \, \uv{y}.
\ee
This velocity, along with a vertical density profile $\rho(z) = \rho_0
\exp[-\Omega^2 z^2/(2c_s^2)]$ and zero magnetic field, is a steady-state
solution to Eqn(\ref{BE2}). $\rho_0$ is the midplane density. In this
work, we nondimensionalize the local model by choosing $\rho_0 = 1$,
$\Omega = 1$, and $c_s = 1$; the usual disk scale height $H$ is
therefore $ H \equiv c_s/\Omega = 1$.  The initial surface density is
therefore $\int \rho dz = \sqrt{2\pi} \rho_o$.

The local model is realized numerically using the ``shearing box"
boundary conditions (e.g. \citealt{hgb95}), which isolates a rectangular
region in the disk.  The azimuthal ($y$) boundaries are periodic; the
radial ($x$) boundaries are ``shearing periodic''; they connect the
radial boundaries in a time-dependent way that enforces the mean shear.
The vertical ($z$) boundaries use a form of outflow boundary conditions:
all variables in ghost zones (including the $z$ velocity and momentum on
vertical boundaries because of the staggered mesh) are copied from the
last active zone in the computational domain, with the additional
constraint that no inflow is allowed.  For stratified disk models the
outflow boundary condition is better motivated than periodic boundary
conditions, although it is more difficult to implement.

What constraint do these boundary conditions place on the field
evolution?  Integrating the induction equation over the computational
domain yields, after application of Stokes theorem,
\be
L_x L_y L_z \del_t \<B_x\> \equiv \del_t \int d^3x B_x = 
\int dx \int d\bld[s] \cdot (\bv \times \bB) =
\ee
where the second integral is taken on a circuit round the box boundaries
at fixed $x$.  It is evident that the EMF integrated over a line on the 
top boundary will not cancel that on the bottom boundary for outflow
boundary conditions, and so $\<B_x\>$ is not conserved.  A similar
argument implies that $\<B_y\>$ is not conserved either.
$\del_t \<B_z\>$ is proportional to a line integral around the box at
constant $z$, where the quasi-periodic radial and periodic azimuthal
boundary conditions do cause cancellation, so $\<B_z\>$ is constant
(numerically: constant to within accumulated roundoff error).

In the preceding paragraph we adopted the notation $\<\,\,\>$ for a
volume average:
\be
\<\,f\,\> \equiv \frac{1}{V} \int_V dx dy dz f.
\ee
We will also use 
\be
[\,f\,] \equiv \frac{1}{L_x L_y} \int dx dy f
\ee
for a plane average, and
\be
\bar{\,f\,} \equiv \frac{1}{T} \int_T dt f
\ee
for a time average.

Our models are evolved using \zeus \citep{sn92} with ``orbital
advection'' \citep[aka FARGO; see]{mass00, gamm01, jg05} for the
magnetic field \citep{jgg08,fs09}.  \zeus is an operator-split, finite
difference scheme on a staggered mesh that uses a Von Neumann-Richtmyer
artificial viscosity to capture shocks (this is a nonlinear bulk
viscosity that does not produce significant angular momentum transport
in our models), and the Method of Characteristics-Constrained Transport
(MOC-CT) scheme to evolve the magnetic field and preserve the $\bnabla
\cdot \bB = 0$ constraint to machine precision.  The orbital advection
is implemented on top of \zeus.  It decomposes the velocity field into a
mean shear part with orbital velocity $\bv_{orb} = -q\Omega x
\hat{\bld[y]}$  and a fluctuating part $\delta \bv$; $\bv = \delta \bv +
\bv_{orb}$.  Advection for the mean flow can is done using interpolation
(which is always stable), so that the Courant limit on the timestep
depends only on $\delta \bv$ and not $\bv_{orb}$.  Shearing boxes with
$L_x \gtrsim H$, where the shear flow is supersonic, can then be evolved
more accurately, and with a larger timestep.

We have also implemented an additional procedure to make the numerical
diffusion more nearly translation invariant in the plane of the disk.
As discussed in \cite{gg09}, the entire box is shifted by a few grid
points in the radial direction at $t = 2 n L_y/(3 \Omega L_x)$, $n =
1,2,3,\ldots$); at these instants the box is exactly periodic.  After
the shift we execute a divergence cleaning procedure to remove the
monopoles that are created by joining the radial boundaries together in
the middle of the computational domain.  This procedure carries little
computational cost. 

The timestep in large stratified disk simulations is limited through the
Courant condition by the Alfv\'en speed $v_A = B/\sqrt{4\pi\rho}$ at
large $|z|/H$, where the density is orders of magnitude smaller than at
$z = 0$.  To prevent the simulation from being brought to a halt by low
density zones (and to avoid other numerical artifacts associated with
small $\rho$), we impose a density floor $\rho _{\rm min} =
10^{-5}\rho_{0}$.  This density floor is $\sim 1-2$ orders of magnitude
smaller than the averaged minimum density in the saturated state.  We
have tested a smaller density floor $\rho_{\rm min} = 10^{-7}\rho_{0}$
and found the choice of the density floor does not affect our
results.

\section{Large Stratified Disk Simulations}

\subsection{Fiducial Model}

All models start from a hydrodynamical equilibrium, with $\rho(z) =
\rho_0 \exp(-z^2/[2H^2])$.  We introduce a uniform toroidal field
$\bB_{0} = B_{0}\hat{\by}$ at $|z| \leq 2H$; $B_{0}$ is chosen so that
at the disk midplane the initial plasma parameter $\beta_{0} \equiv 8\pi
P_0/B_0^2 = 25$ (the sharp vertical variation in $B_y$ at $|z| = 2 H$
makes the disk initially unstable to magnetic Rayleigh-Taylor
instability, but this structure is quickly wiped out by MRI driven
turbulence).  Each component of the velocity is perturbed in each
zone, with $\delta v_i$ uniformly distributed in $[-0.01,0.01] c_s$.
The models are evolved long enough ($\geq 150$ orbits $\sim
900\Omega^{-1}$) to reach a saturated, i.e., statistically steady,
state.

\begin{figure}
\plotone{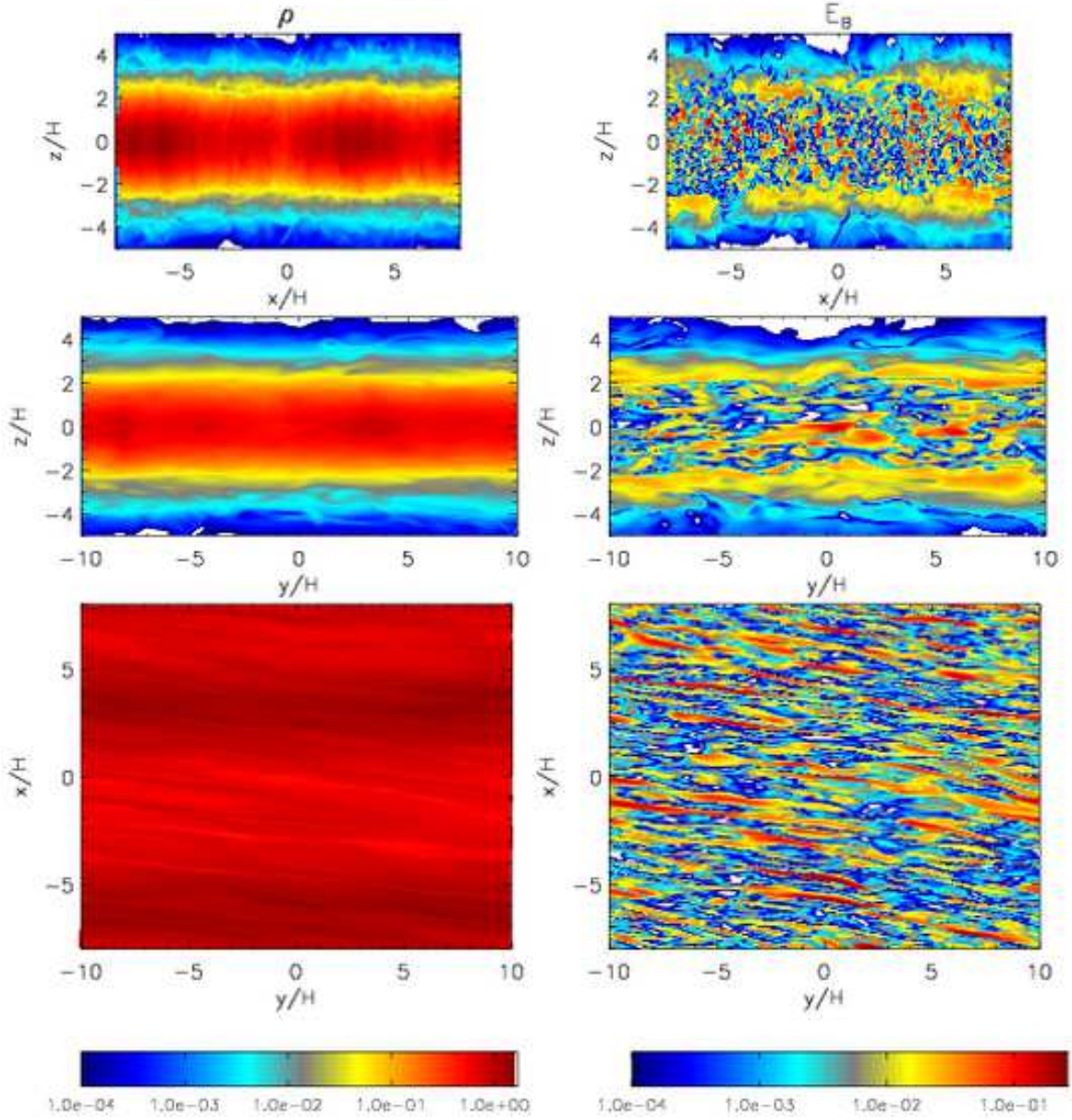}
\caption{Snapshots of density and magnetic energy density in the
  fiducial model, taken at $t = 100$ orbits. Left: density $\rho$; right: magnetic energy density $E_B$;  top: image at $y = 0$ plane; middle: image at $x = 0$ plane; bottom: image at $z=0$ plane. 
}
\label{fig:xz.images}
\end{figure}

Our fiducial model has a domain size of $(L_x, L_y, L_z) = (16, 20,
10)H$ and resolution $384\times 256\times 128$.  This corresponds to a
physical resolution of $(24, 12.8, 12.8)$ zones per scale height.
Snapshots of $\rho$ and $E_B \equiv B^2/8\pi$ at slices with constant
$x,y$ and $z$ in the saturated state are shown in Figure
\ref{fig:xz.images}.  

Turbulence is confined to the region $|z| \leq 2H$.  Within this region
magnetic field fluctuations are contained on a scale $l \ll H$, with a
structure in the shape of narrow filaments that are extended by the
azimuthal shear.  This turbulent field structure resembles that observed
in unstratified disk simulations \citep{gg09}. Density fluctuations on a
scale $\sim H$ due to sound waves are also evident in the $x-z$ plane
density snapshots. At $|z| > 2H$ the MRI is suppressed.  $E_B$ decreases
sharply, but not as rapidly as $\rho$.

\begin{figure}
\plotone{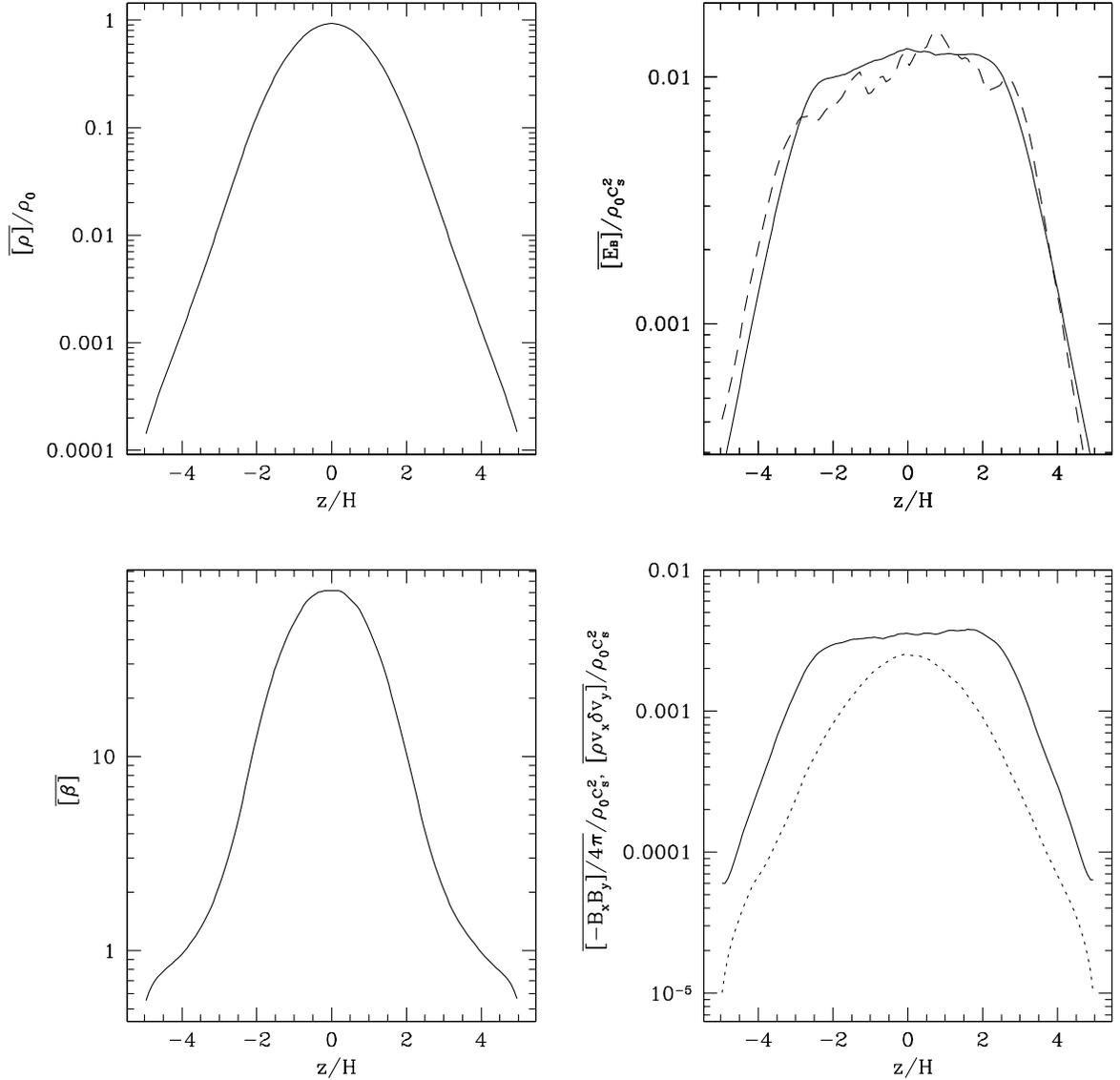}
\caption{
Vertical profiles of several $x-y$ plane averaged quantities in the fiducial
model.  Upper left: density;
Upper right: magnetic energy density; Lower left: plasma $\beta$; Lower right: $xy$
component of Maxwell stress $M_{xy} = -B_xB_y/4\pi$ (solid lines) and
Reynolds stress $R_{xy} = \rho v_x\delta v_y$ (dotted lines). All
quantities in solid and dotted lines are averaged from the last $50$ orbits. To
illustrate the time average effect, we also plot $[E_B](z)$ at $t =
900\Omega^{-1}$ (dashed lines) in the upper right panel. The slight asymmetry of $\overline{[E_B]}(z)$ and $\overline{[M_{xy}]}(z)$ in the
$|z| \leq 2H$ region is probably due to our choice of orbits interval
for time average.
}
\label{fig:f.vs.z}
\end{figure}

The disk vertical structure is shown in Figure \ref{fig:f.vs.z}, which shows $\overline{[\rho]}$, $\overline{[E_B]}$,
$\overline{[\beta]}$, Maxwell (magnetic) stress $\overline{[M_{xy}]}
\equiv \overline{[-B_x B_y]}/4\pi$ and Reynolds stress
$\overline{[R_{xy}]} = \overline{[\rho v_x\delta v_y]}$.  These profiles
are obtained from a time average over the last $100\pi/\Omega$. The most 
striking feature in these profiles
is the ``turbulent disk surface'' at $|z| \sim 2H $ defined by
$\overline{[E_B]}(z)$ and $\overline{[M_{xy}]}(z)$. Inside this surface
both are independent of $z$; outside both exhibit an approximately
exponential dependence on $z$.  As illustrated in the vertical profile
of $\beta$, as $|z|$ increases, magnetic energy density drops slower
than density; above $|z| \sim 2.5H$ $\beta$ drops below unity. Therefore
the region $|z| > 2H$ is magnetically dominated and this leads to the
suppression of the MRI.  From now on we will simply refer to the
magnetically dominated upper region with $\beta < 1$ as ``corona",  and
the turbulent $|z|\leq 2H$ region as ''disk.''

Fits to the disk structure give
\begin{equation}
\overline{[\rho]} (z) \simeq \cases{
0.93\rho_o\,\exp(-\frac{z^2}{2H}), & if $|z| \leq 2.55H$; \cr
0.036\rho_o\, \exp{(-\frac{|z|-2.6H}{0.44H})}, & otherwise.\cr}
\label{eqn:rhoz}
\end{equation}
and
\begin{equation}
\overline{[E_B]}(z) \simeq \cases{
  0.012\rho_0 c_s^2, & if $|z| \leq 2.55H$; \cr
  0.012\exp{(-\frac{|z|-2.6H}{0.64H})}\rho_0 c_s^2, & otherwise.\cr}
\label{eqn:ebz}
\end{equation}
In the saturated state, $\overline{[\rho]} (z)$ is different from the
initial density profile $\rho_0\exp[{-z^2/(2H^2)}]$ due to the magnetic
buoyancy effects and mass loss through the $z$ boundaries. Inside the
disk, a nearly Gaussian density profile indicates that this region is
still mainly supported by gas pressure.

How can we understand the vertical magnetic structure of the disk?  A
uniformly magnetized atmosphere is subject to interchange and Parker
type modes \citep{newcomb, parker66}.  The more dangerous of these is
Parker, whose stability condition is
\begin{equation}  
-\frac{d\rho}{dz} > \frac{\rho^2 g}{\gamma P_{\rm gas}}  ,
\label{eqn:parker}
\end{equation}
where $P_{\rm gas}$ is the gas pressure, $g = \Omega^2z$ is the
gravitational acceleration, and $\gamma$ is the adiabatic index (here,
$\gamma = 1$) \citep{newcomb}.  For a disk in hydrostatic equilibrium  
\begin{equation}
-\frac{d(P_{\rm gas} + P_{\rm mag})}{dz} = \rho g;
\label{eqn:hydro}
\end{equation}
together these conditions imply
\begin{equation}
\frac{dP_{\rm mag}}{dz} = \frac{dE_B}{dz} = 0.
\end{equation}
Marginally stable stratification therefore corresponds to constant
$\overline{[E_B]}$, as is found at $|z| < 2 H$.  This suggests that (1)
magnetic buoyancy is driving the disk toward a marginally stable state;
(2) magnetic buoyancy is crucial in controlling the vertical magnetic
structure in the bulk of the disk.  If this is correct, it follows that
$E_B(z)$ in the disk could be different in nonisothermal models.  In
particular, marginal stability requires
\be
\frac{1}{8\pi}\frac{d [B^2]}{dz} = \gamma P_{gas} \left(\frac{1}{\gamma}\frac{d\ln P_{\rm gas}}{dz} -
\frac{d\ln \rho}{dz} \right).
\ee
Thus an isentropic disk has $d B^2/dz = 0$, while a stably stratified,
nonradiative disk (in the Schwarzschild sense) can support $d B^2/dz <
0$.  In a radiative disk the instability criterion is modified (disks
heated by internal dissipation of turbulence rather than external
irradiation tend to have strong radiative diffusion, or Peclet numbers of
order $\alpha^{-1} \sim 50$), because radial radiative diffusion tends
to wipe out temperature perturbations for the most unstable modes with 
high radial wavenumbers.

\begin{figure}
\plotone{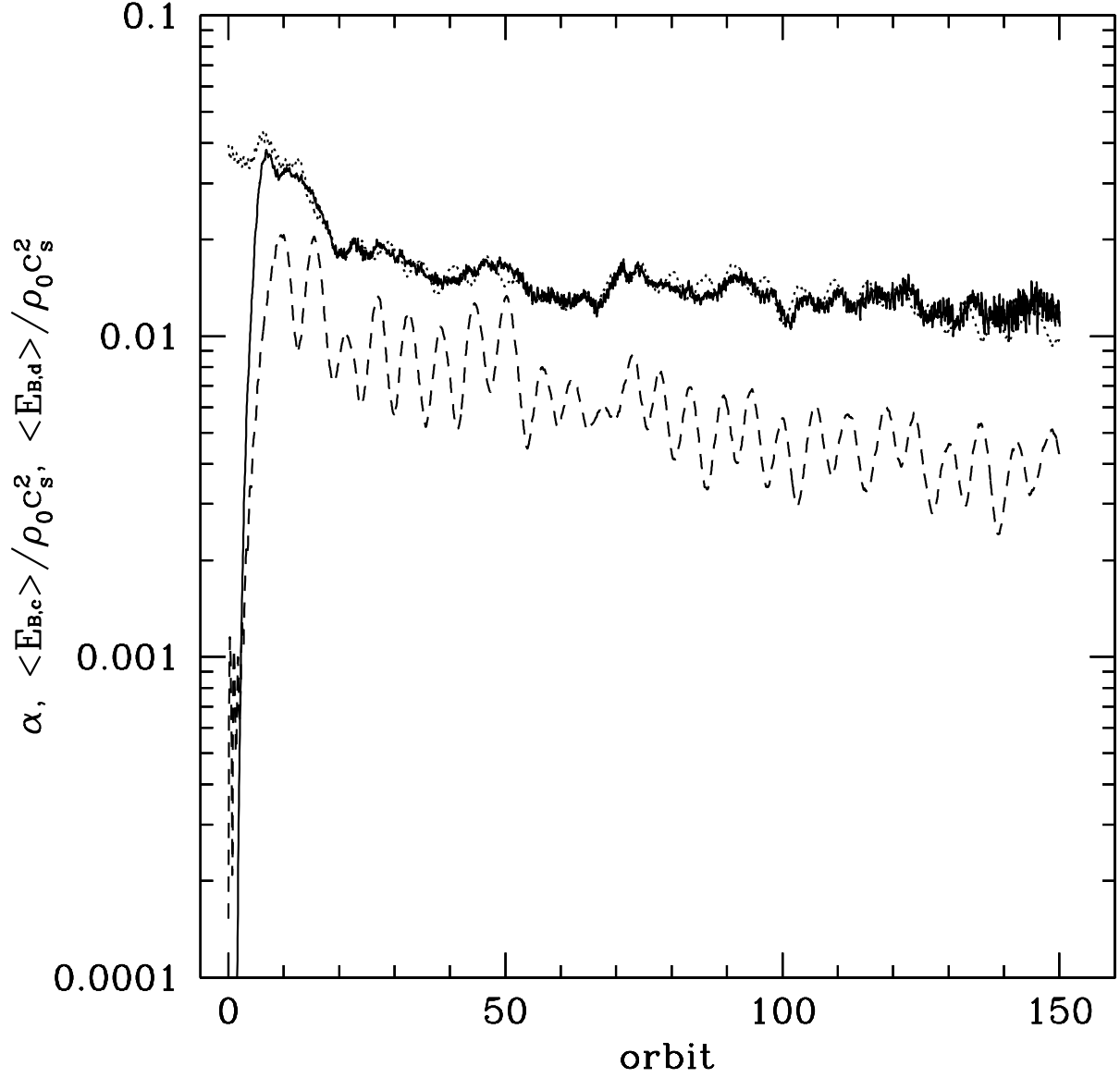}
\caption{Evolution of $\<\alpha\>$ (solid lines), $\<E_{B,d}\>/\rho_0c_s^2$ at $|z|
  \leq 2H$ (dotted lines), and $\<E_{B,c}\>/\rho_0c_s^2$ at $|z| > 2H$
  (dashed lines) in the fiducial model. Saturation $\overline{\<\alpha\>} \sim 0.013$ when
averaged over the last $50$ orbits.
}
\label{fig:f.vs.t}
\end{figure}

Figure \ref{fig:f.vs.t} shows the evolution of magnetic energy density
in the disk $\<E_{B, {\rm d}}\>$, magnetic energy density in the corona
$\<E_{B, {\rm c}}\>$ and $\<\alpha\>$ 
\begin{equation} 
\<\alpha\> \equiv \frac{\int W_{xy} d^3x}{\int \rho c_s^2 d^3x} 
\end{equation} 
where $W_{xy} \equiv R_{xy} + M_{xy}$ is the total shear stress
Averaging the last $50$ orbits, we found that $\overline{\<\alpha\>}
\sim 0.013 $, $\overline{\<E_{B, {\rm d}}\>}/\rho _{0}c_s^2 \sim 0.012$,
$\overline{\<E_{B, {\rm c}}\>}/\rho _{0}c_s^2 \sim 0.0043$. 

\subsection{Two-point correlation function}

One question motivating this study was whether thin disks exhibit
mesoscale structure, i.e. structure on scales that are $\gg H$ but $\ll
R$.  As is evident in Figure \ref{fig:xz.images}, the characteristic
scale of the magnetic energy density varies with $|z|$.  Near $|z| = 0$,
turbulent structure resembles that observed in unstratified disk models:
the field is confined in small structures with scale $l \ll H$.   Away
from $|z| = 0$, $l$ increases, reaching $\sim H$ at $|z| \sim 2.5 H$.

\begin{figure}
\plotone{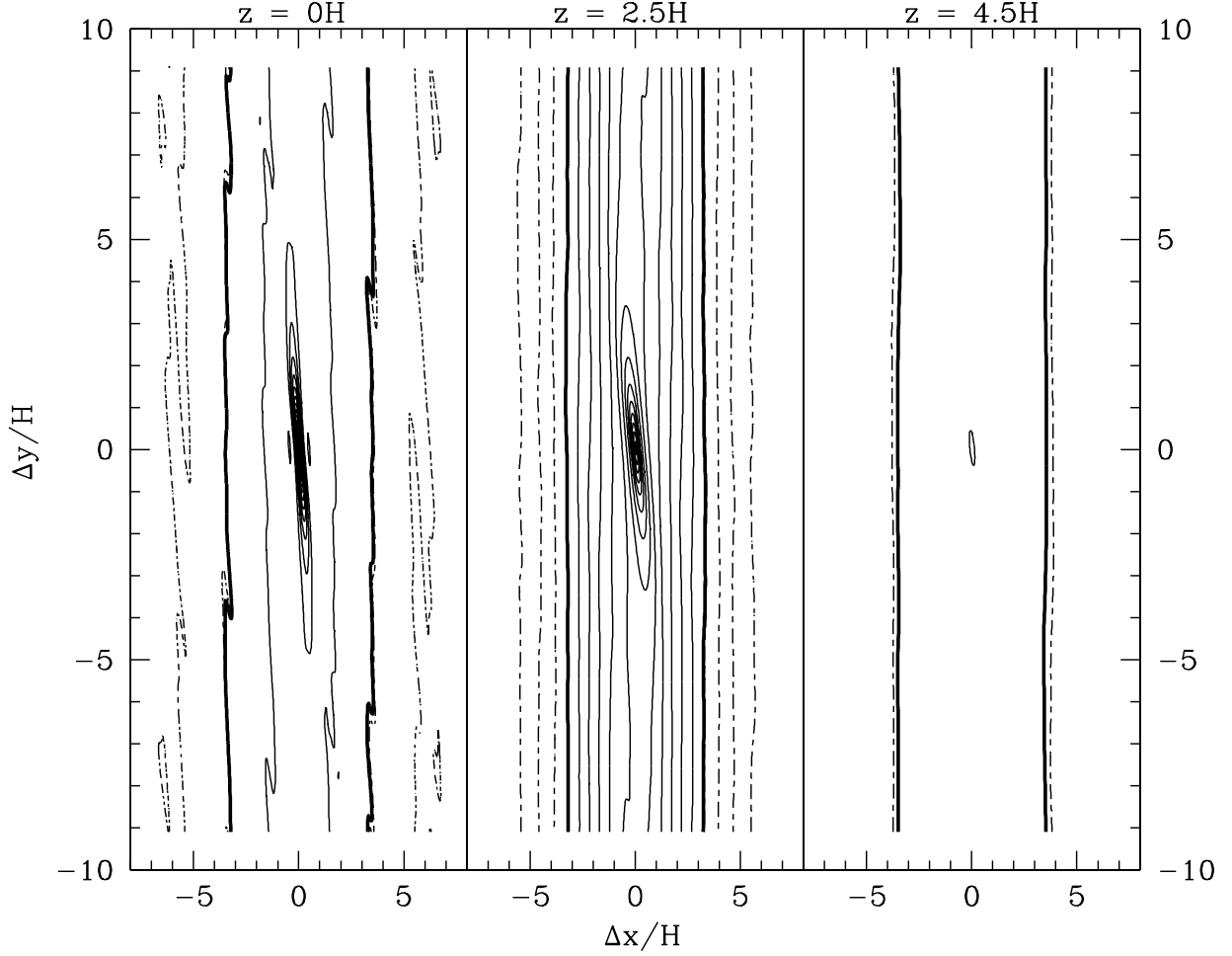}
\caption{ Contour plots of 2D two-point correlation function $\xi_B(z)$ for $\delta\bB$. Plotted are $\xi_B(z)/(4\pi
\rho_0c_s^2)$ in the $(\Delta x, \Delta y) $ plane at three different
vertical heights in the fiducial model. Left: mid plane; Middle: $z = 2.5H$;
Right: $z = 4.5H$. The contours run linearly from $-0.058$ to $0.229$ for
$20$ levels; solid lines:  $\xi_B \geq 0$; dash lines: $\xi_B < 0$;
the heavy line is the $0$ contour.
}
\label{fig:corr.xy}
\end{figure}

The two-point correlation function $\xi$ provides a quantitative measure
of disk structure:
\begin{equation}
\label{xi}
\xi_B (z) \equiv [\delta \bB(x, y; z) \cdot \delta \bB(x + \Delta x,
y+\Delta y; z) ].
\end{equation}
Here $\delta \bB \equiv \bB - [\bB]$; for a detailed discussion of $\xi$
and the corresponding correlation lengths $\lambda_i$ see
\citep{ggsj09}.  Figure \ref{fig:corr.xy} shows $\xi_B(z)$ in the
$(\Delta x, \Delta y)$ plane at $z = 0$,  $z = 2.5H$ and $z = 4.5H$.  In
these plots we have averaged 8 neighboring vertical zones to increase
the signal-to-noise ratio. At the disk midplane, the correlation
function has a narrow elliptical core of width $\lambda < $ a few $H$.
As $|z|$ increase to $2.5 H$ the core becomes larger, especially in the
radiation direction, and low amplitude features develop on scales of
$\sim 10 H$.  These low-amplitude, mesoscale features are new and are
not seen in unstratified disk models.

\subsection{Coronal loop structure}

Our disk models contain a ``corona'', where $\beta < 1$.   It is not
clear how accurately, or inaccurately, our code models this region
because it contains no explicit model for reconnection (nor is any
convincing model currently available; see \cite{uzgoo} for a
discussion of the difficulties of simulating force-free
coronae). Still, it is interesting to characterize the field structure
in existing simulations before asking how they might be changed by more sophisticated reconnection models.

\begin{figure}
\plotone{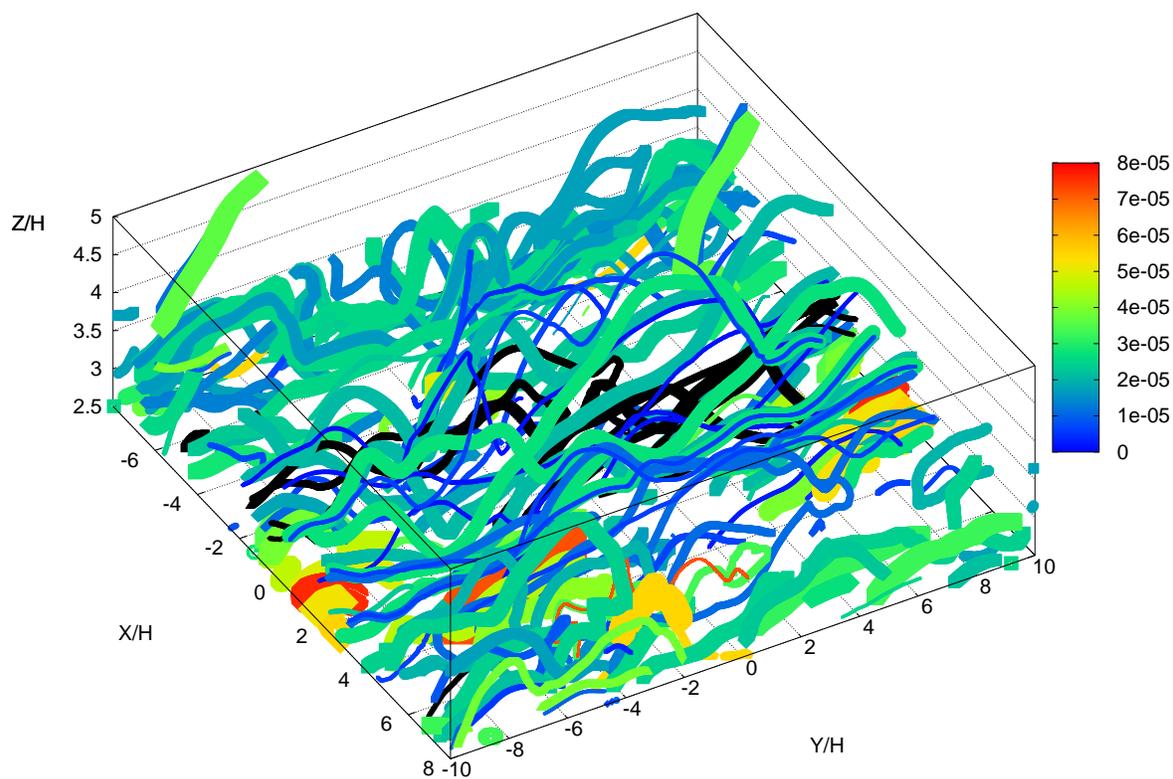}
\caption{ Magnetic field lines originating from the plane $z = 2.5H$ in
  the fiducial model at $t = 600\Omega ^{-1}$. The lines are evenly
  sampled spatially from the $x-y$ plane. Both the line width and the color (see the online version for a color version of this plot) denote the flux carried by each line at the footpoint, normalized by the total flux from the $z =
  2.5H$ plane. Majority of the
  field lines return to $z = 2.5H$, forming closed loops. 
}
\label{fig:loops}
\end{figure}

How can we understand coronal magnetic field structure?  Most of the
coronal field is anchored in the disk, so we begin by sampling field
lines that rise through the surface $z = 2.5 H$ at a single
instant. Using bilinear interpolation for the field, we trace field
lines initiated from every cell on the $z = 2.5H$ surface, until they
either (a) come back to the $z=2.5H$ surface, or (b) leave the upper
$z$ surface, or (c) exceed maxmum integration step $10^5$ indicating
the formation of a closed loop.  A snapshot of these field lines are show in Figure \ref{fig:loops}. Two features of the coronal field are obvious just
from visual inspection: many of the field lines return to the disk after
only a short sojourn in the corona, and the loops tend to have greater
azimuthal than radial extent.

A more quantitative approach is to calculate a coronal loop distribution
function, as in the phenomenological model of \cite{uzgoo} (hereafter
UG).  The field lines should then be sampled according to the flux
through each zone surface $d\Phi_i = B_{z,i} dx dy$. We also average
over the last 50 orbits to improve the loop statistics. We find that
$\sim 96\%$ of the field lines passing through the $z=2.5H$ surface
return to the same surface, $\sim 4\%$ of field lines are open in the
sense that the escape through the upper boundary of the box, and only
$\sim 0.1\%$ of the field lines form closed loops inside the
corona. We have also found that the small fraction of the
open field lines is quite stable during the saturated state, ranging
from $\sim 2 - 5 \%$ at instaneous state, therefore it appears that
the corona fields structure has reached a statistical steady state.   

We then use three variables to describe the geometry of close field
lines (field loops) that return to the $z = 2.5 H$ surface: the loop
foot point separation $\Delta \bx$ in the $x-y$ plane, the loop maximum
height $\Delta z_{\rm max}$, and the loop orientation angle $\theta
_{\rm foot} \equiv$ the angle between the foot separation vector and the
$y$ axis.  We calculate the distributions functions ${d\Phi}/{d\Delta
x}$, ${d\Phi}/{d\Delta y}$, ${d\Phi}/{d\Delta z_{\rm max}}$ and
${d\Phi}/{d\theta_{\rm foot}}$ by following the trajectory of every
field line that emerges at the center of each zone surface $i$, then
weighting the result by the flux $d\Phi_i$.  The final distribution
function is normalized by $|\Phi|$, the total absolute flux through the
$z = 2.5 H$ plane. 

\begin{figure}
\plotone{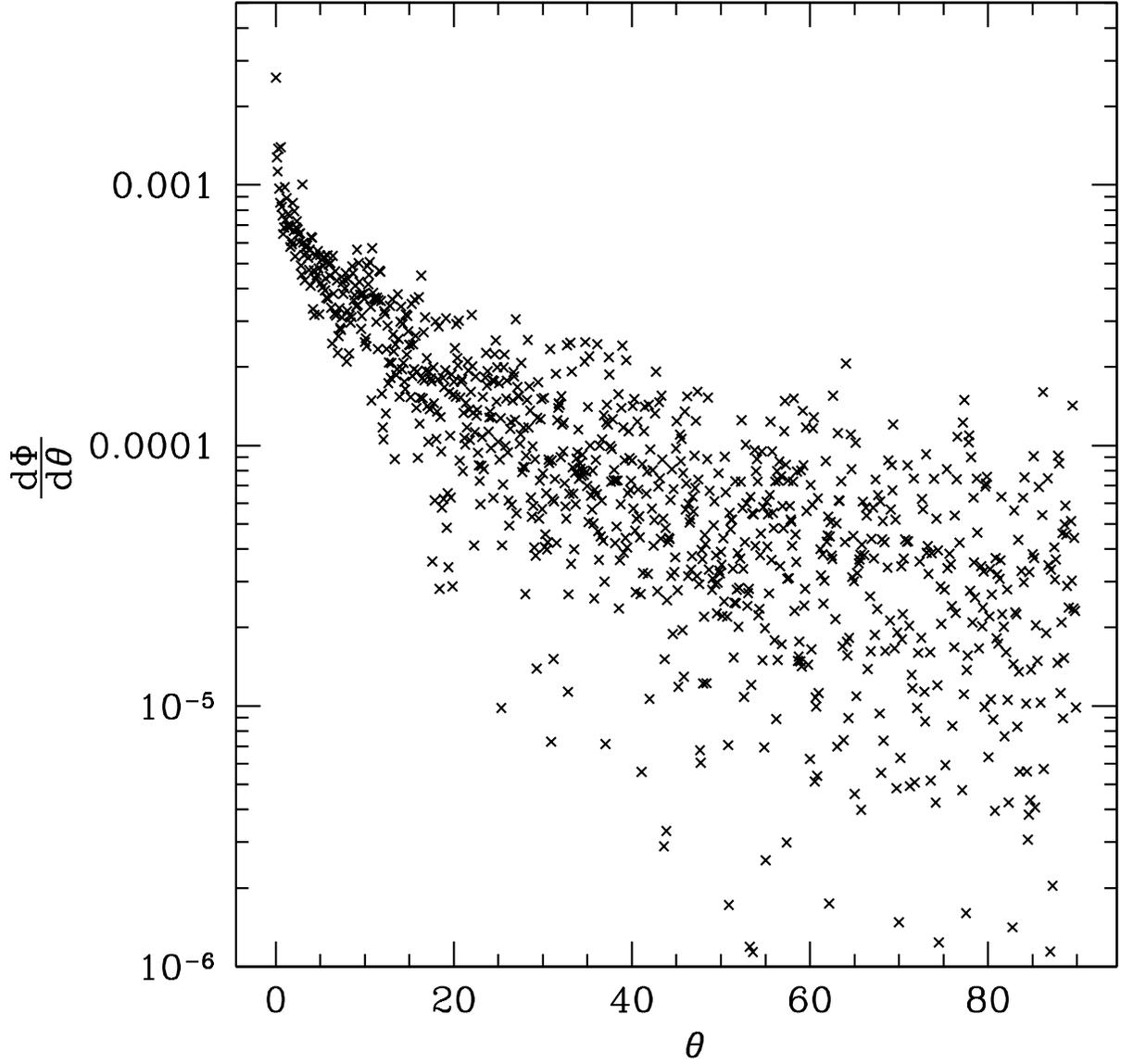}
\caption{ Loop angle distribution functions in the fiducial model:
  $\theta$ denotes the angle between the foot separation vector
  and the $y$ axis. Loops are stretched in the azimuthal direction due
  to the shear. 
}
\label{fig:looptheta}
\end{figure}

\begin{figure}
\plottwo{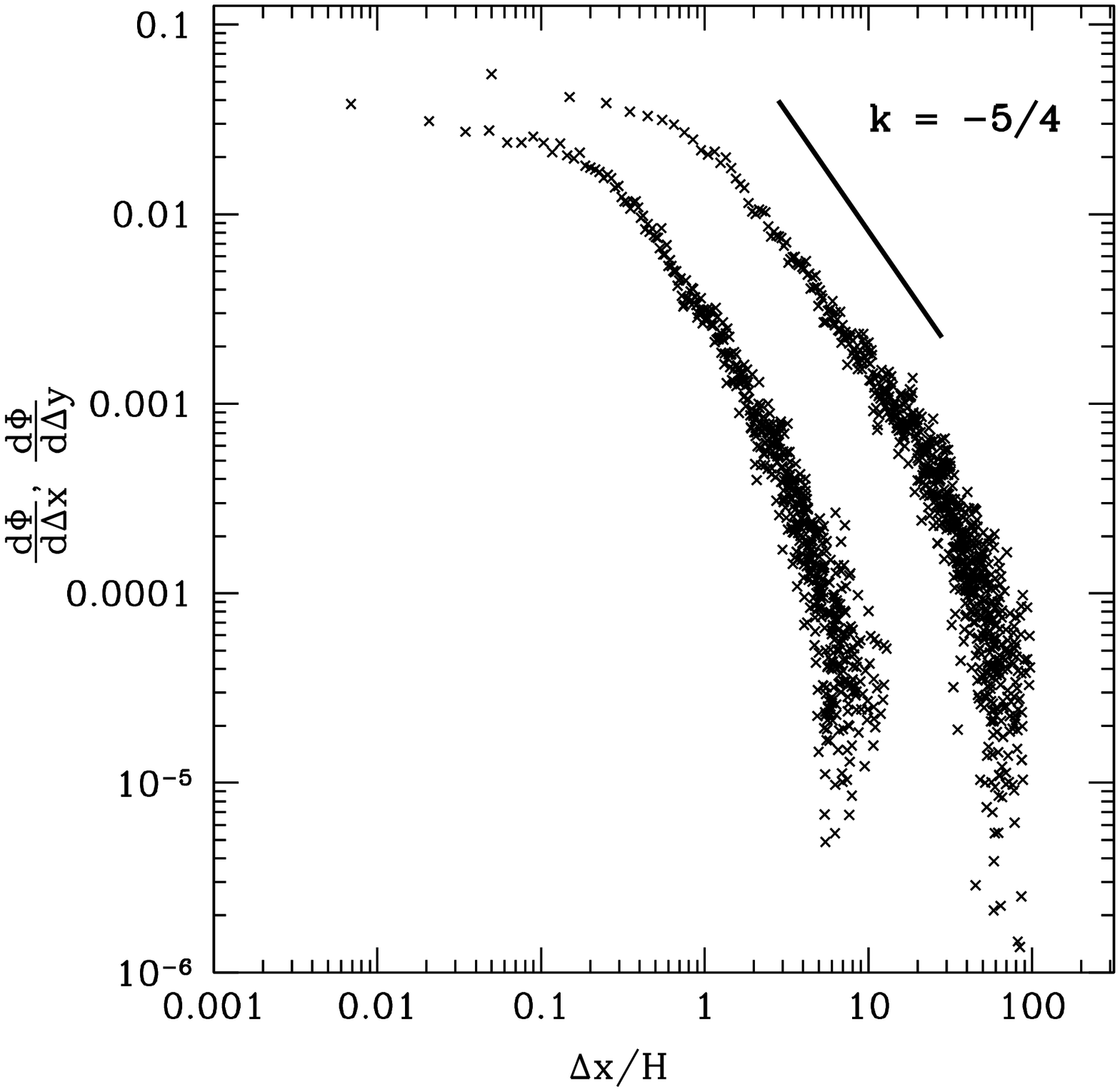}{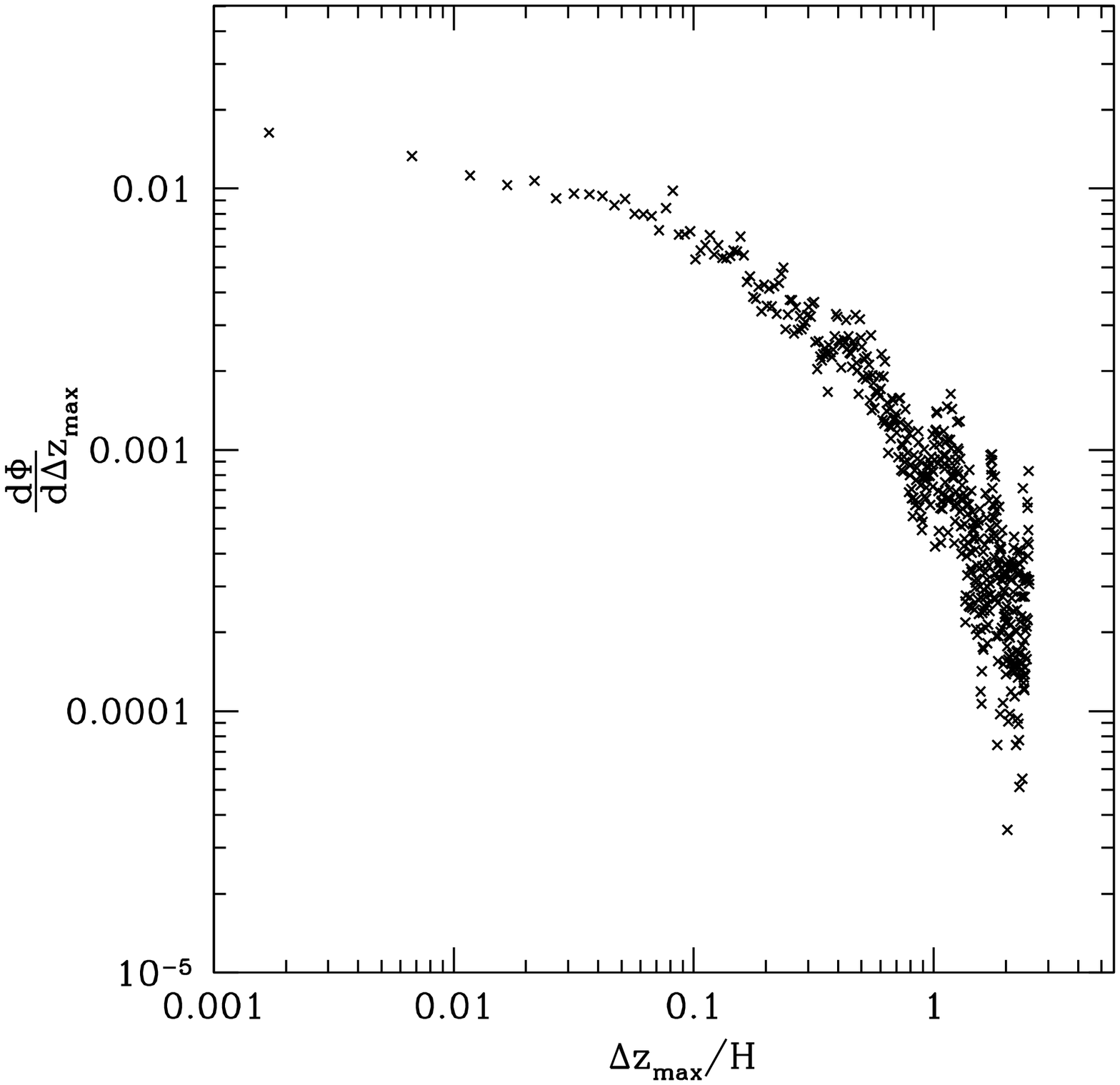}
\caption{ Loop-distribution functions in the fiducial model.
 Left panel: foot point separation distribution. Left curve is for $\Delta
 x$ and  Right curve is for $\Delta y$. The distribution for $\Delta L
 = (\Delta x^2 + \Delta y^2)^{1/2}$ almost overlaps with the $\Delta
 y$ curve. The heavy line indicates a $k = -5/4$ slope. Right panel:
 loop height distribution. 
}
\label{fig:loopdist}
\end{figure}

Figures \ref{fig:looptheta} and \ref{fig:loopdist} show
${d\Phi}/{d\Delta x}$, ${d\Phi}/{d\Delta y}$, ${d\Phi}/{d \Delta z_{\rm
max}}$ and ${d\Phi}/{d\theta _{\rm foot}}$, averaged over the last $50$
orbits.  From ${d\Phi}/{d\theta_{\rm foot}}$ (Figure
\ref{fig:looptheta}) it is evident that most of the field loops are
orientated in the azimuthal direction with $\theta \leq$ a few degrees.
This suggests that shear plays a significant role in determining the
coronal field structures.  Most loops also have maximum height $\Delta
z_{\rm max} \leq H$ (see ${d\Phi}/{d\theta _{\rm foot}}$ in Figure
\ref{fig:loopdist}). 

If there is no reconnection at all then magnetic energy injection from
the underlying turbulent disk might cause the loop to grow in an
unlimited way. This is not the case here: although we do not include
dissipation explicitly, numerical reconnection due to truncation errors
is present in our numerical scheme, as in all finite-difference MHD
schemes.  It is difficult, however, to quantify the numerical
reconnection rate in our model directly.  We therefore try to compare
our numerical loop distributions to the predictions of the UG model.

We fit a power-law to ${d\Phi}/{d\Delta x}$ and  ${d\Phi}/{d\Delta y}$,
\begin{equation}
\frac{d\Phi}{d\Delta x} \simeq C_0 (\frac{\Delta x}{H})^{k}\,\,,
\end{equation}
where $C_0$ is a constant. Notice the shape for ${d\Phi}/{d\Delta x}$ and ${d\Phi}/{d\Delta y}$ are very similar. For $\Delta y$ a fit between
$3H\leq \Delta y\leq 20H$ gives $k \approx -1.2$. We can also calculate the general loop distribution function for $\Delta L = (\Delta x^2 + \Delta y^2)^{1/2}$. It almost overlaps with the ${d\Phi}/{d\Delta y}$ curve because the loops
are nearly toroidal, and a fit between $3H\leq \Delta y\leq 20H$ gives $k \approx -1.2$.

In the model of UG, a slope of $k\sim -2$ corresponds to the limit that
reconnection is slow compared to the shear (the dimensionless
reconnection parameter $\kappa \sim 0.01$ in UG), and a slope of $k >
-1.5$ corresponds to the cases when the total magnetic energy of the
corona is dominated by the largest loops ($\kappa < 0.002$). The shallow
$k \sim -1.2$ slope measured here then indicate our numerical models are
probably in a slow Sweet-Parker reconnection regime. However, this
comparison should not be taken too seriously\footnote{Although the
reconnection rate in the corona might well determine the vertical
magnetic energy profile $[E_{B,c}(z)]$, simply from a characteristic
field curvature argument, where $ l(z) \sim [ v_a(z)]/\Omega \sim \Delta
\bx(z)$. } because our model is ideal MHD, and does not explicitly model
reconnection. One serious concern is that the coronal reconnection could
fall into a fast, collisionless regime which is poorly understood, and
not well modeled by our grid scale dissipation.

Lastly we want to comment on several surface effects in our model. These
include the $yz$ component of magnetic stress tensor $M_{yz} \equiv -B_y
B_z/(4\pi)$, the vertical components of kinetic flux and Poynting flux,
and the mass loss rate. Notice these quantities do not necessarily
average to zero because of the outflow boundary conditions. We have
found that $\<M_{yz}\>$ is nearly zero with temporal fluctuations of
amplitude $\leq 10^{-4}\rho_{0} c_s^2$, much smaller than the dominant
$xy$ component $\<M_{xy}\>$.  In the steady state, the vertical energy
flux is dominated by the advective part of the Poynting flux, which is
on the order of $10^{-4}\rho_0 c_s^3$. This vertical energy flux is only
$\sim 1\%$ of the turbulent dissipation rate $Q$ in the disk main body
($Q \sim \alpha \rho_{0} c_s^3 \sim 10^{-2} \rho_{0} c_s^3$), indicating
a weak vertical energy flux.

The disk loses mass through the upper and lower boundaries. In the last
$50$ orbits, the disk lost $\sim 1.4\%$ of its initial mass.  The
vertical mass loss rate is not negligible\footnote{The ratio of vertical
mass loss rate to the mass accretion rate is
\begin{equation}
\frac{\dot{M}_z}{\dot{M}_r} \sim \frac{\pi r^2\frac{\delta
    \Sigma}{\delta t}}{3\pi\Sigma\nu} = \frac{\pi
  r^2\frac{\delta\Sigma}{\delta t}}{3\pi\Sigma\frac{\overline{\alpha}
    c_s^2}{\Omega}} = \frac{1}{3} \frac{\delta
  \Sigma}{\Sigma}\frac{1}{\overline{\alpha}}({\frac{r}{H}})^{2}\frac{1}{\delta
t\Omega}\sim 0.001(\frac{r}{H})^{2},
\end{equation}
where $\dot{M}_r$ is the mass accretion rate at the disk radius $r$,
$\delta \Sigma$ is the change of disk surface density in $\delta t$, and we
have used a disk turbulent viscosity $\nu = \alpha c_s^2/\Omega$. For
$r/H = 30$, $\dot{M}_z/\dot{M}_r \sim 1$.} 
in our $L_z = 10H$ models because of the outflow boundary conditions.
However, we have noted a trend in which the vertical mass loss decreases
with increasing $L_z$. For example, in our $L_z = 12H$ model the disk
lost $\sim 0.64\%$ of its initial mass during the last $50$ orbits,
giving a mass loss rate half that of the $L_z = 10H$ model. We therefore
expect a decreasing mass loss rate as $L_z$ increases. 

It is also worth noting that in our models the mean vertical magnetic
fields maintain $\<B_z\> = 0 $ because of the shearing box boundary
conditions. We also found that the plane-averaged $[B_z] \sim 0 $ at all
$z$ including at the domain boundaries. The vertical field at the
surfaces are turbulent with patches of opposite sign field penetrating
the boundaries.  However the vertical field here are fluctuations with
radial correlation length $<$ a few $H$ and amplitude $\sim$ one order
of magnitude smaller than that at the disk mid plane. We have not
observed a steady magnetic wind and the observed mass loss is probably
due to outflow boundaries.

To summarize, in our models we see a weak wind launched from the disk
surface. In a steady state both vertical energy and momentum flux are
negligible.

\subsection{Dependence on Model Parameters}

Here we give a brief discussion of the saturation dependence on model
parameters,  including: (1) resolution; (2) $L_x$;  (3) $L_z$; (4)
initial field strength in terms of plasma parameter $\beta_{0}$. When
exploring parameter space we vary only one parameter at a time unless
stated otherwise. Model parameters can be found in Table 1.

(1) Resolution. In model s16a, we test the convergence properties of our
numerical models by doubling the resolution of the fiducial run to
$768\times 512\times 256$. We run this model to $t_f \sim 75$ orbits.
Averaging over the last $\sim 25$ orbits in the satuated state, we found
that $\overline{\<\alpha\>} \sim 0.023$, almost double of that in the
fiducial run ($\overline{\<\alpha\>} \sim 0.013$); at the highest
resolution explored in this work, saturation level continues to increase
with resolution. The dynamic range in resolution explored in this work
is modest (highest resolution in this work is $\sim 20-40$ zones per
$H$) due to the computational demands of the large box\footnote{For
example, a $(L_x, L_y, L_z) = (16, 20, 10)H$ run with resolution
$768\times 512\times 256$ (run s16a) and $t_{\rm f} \sim 150$ orbits
required $\sim 0.5\times 10^6$ cpu hours on \abe\ cluster at NCSA.}. We
have also monitored ``quality factor '' $Q$, where $Q_i \equiv
\lambda_{\rm MRI, i}/\Delta x_i = 2\pi v_{\rm a,i}/(\Omega \Delta x_i) $
(see a discussion of $Q$ in \cite{nkh10}), the zones per most unstable
linear MRI wavelength in our calculations.  For the bulk of the disk
inside $\pm 2H$ region, our highest resolution run gives a volume and
time averaged $Q_y = 33.7$ and $Q_z = 6.8$ in the saturated state, and
so by this measure the toroidal field MRI is well-resolved while the
vertical field MRI is marginally resolved.  Of course, the evolution of
the disk is not well described by linear theory in the fully turbulent
state, so it is not clear whether $Q$ is a good indicator of when
MRI-driven turbulence is sufficiently resolved.

It is worth mentioning the convergence properties of shearing box
simulations done in smaller boxes: (a) the unstratified box with a net
toroidal field, (b) the stratified box with periodic vertical boundary
conditions and (c) the stratified box with outflow boundaries.  First,
using a similar algorithm in unstratified disk simulations with a mean
toroidal field, \cite{ggsj09} reported that in the resolution range
$32-256/H$ saturation energy increases with resolution ($\propto
N_x^{1/3}$). They also pointed out that convergence is expected at
higher resolution when the energy containing eddies are resolved. For
the stratified disks, \citep{skh10} used $L_x=2H$ stratified shearing
boxes simulations with vertical outflow boundaries and they also found
that $\overline{\<\alpha\>}$ increases with resolution and an
$\overline{\<\alpha\>}\sim 0.035$ in their highest resolution run at
$32/H$.   Recently, stratified disk simulations done in smaller boxes
($L_x \sim H$) and periodic boundary conditions with zero-net flux have
demonstrated convergence with $\overline{\<\alpha\>} \sim 0.01$ with a
resolution $\sim 32-128/H$ using \athenap\ code and periodic boundary
conditions \citep{dsp10}. The sustained turbulence may be due to the
presence of a mean toroidal field in the disk midplane.  Notice that
$\overline{\<\alpha\>}$ in their work is normalized with the initial
midplane pressure $P_0$, which is normally a factor of a few larger than
domain-averaged $\<P\>$ used here. Using the definition in this work,
their$\overline{\<\alpha\>} \sim 0.04$. It is therefore possible that in
stratified disk simulations, net-toroidal-field and zero-net-flux models
will have similar convergence properties. If this is the case, we then
expect a convergence at $64-256/H$ using our \zeusp-type code\footnote{A
run of our fiducial run size with a resolution $64/H$ would require
$>5\times 10^6$ cpu hours on NCSA's \abe\ cluster.}.

(2) $L_x$. For $L_x$, we have carried out runs with $L_x = H, 8H$, and
$32H$, denoted by s1, s8, and s32. Time averaging the last $50$ orbits
in each run, we found the saturation level in all these runs are close,
with $\overline{\<\alpha\>} \sim 0.0191\pm 0.00453$ when $L_x = H$,
$\overline{\<\alpha\>} \sim 0.0124\pm 0.00116$ when $L_x = 8H$,
$\overline{\<\alpha\>} \sim 0.0125\pm 0.000965$ when $L_x = 16H$, and
$\overline{\<\alpha\>} \sim 0.0269\pm 0.00211$ when $L_x = 32H$, where
the numbers after $\pm$ denotes standard deviation $\sigma$. The
dependence of $\overline{\<\alpha\>}$  on the box size is not clear,
however, it is difficult to measure $\overline{\<\alpha\>}$ in the $L_x
= H$ box because of the large fluctuations.  In runs with $L_x \geq 8H$,
the $\sigma$-to-mean ratio is aroud $0.07-0.09$, while $L_x = H$ gives a
$\sigma$-to-mean ratio $\sim 0.25$.  

\begin{figure}
\plotone{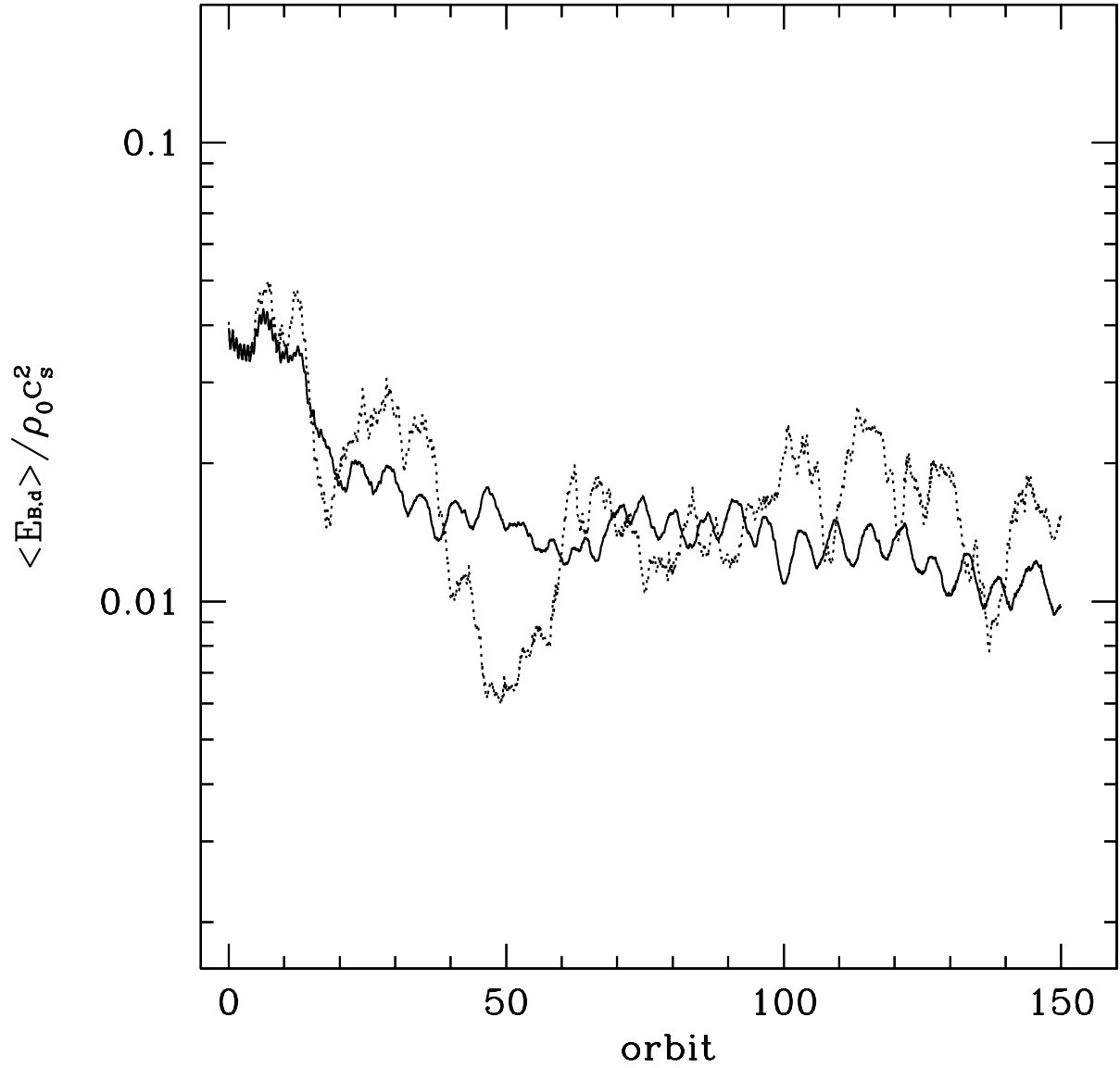}
\caption{ Evolution of the disk magnetic energy density $\<E_{B,d}\>$ in the model
  std16 (solid lines) and s1 (dotted lines). Plotted are the mean magnetic energy density in the region
$|z| \leq 2H$. }
\label{fig:leb.cmp}
\end{figure}

Past stratified disk studies (\citep{dsp10,skh10}) have shown that there
exist significant (order of unity) long term fluctuations in $L_x \sim
H$ box.  The evolution of magnetic energy density in the disk
$\<E_{B,d}\>$ for $L_x = H$ and $L_x = 16H$ runs are shown in Figure
\ref{fig:leb.cmp}.  The smaller fluctuation in large $L_x$ models
suggest that: (a) parts of the disk with horizontal separation $>H$ are
uncorrelated, and (b) the volume integration over large-enough domain
will smooth out these local fluctuations. 

What have we learned in these large domain size models with $L_x \geq
10H $? Our $L_x = H$ run is similar to the toroidal model of
\cite{ms00}. First, in large box runs, the plane-averaged vertical disk
structures are similar to those in smaller box runs: we have observed a
gas-pressure supported disk with a Gaussian density profile inside $\sim
2H$ and an extended magnetic dominated corona outside $\sim 2H$. Second,
the long-term average of disk turbulence saturation level is also very
similar to the $\sim H$ runs, albeit with much smaller temporal
fluctuations.  Statistically, for saturation measurement purposes, a
large domain run can be regarded as a sum of smaller $H$ run, where the
temporal and spatial fluctuations are smoothed out by integrating over
decorrelated disk regions. 

Our models also suggest that a magnetically dominated corona cannot be
studied in an $L_x \sim H$ box (if it can be studied in a numerical MHD
model at all). In large domain size runs with $L_x \geq 10H$ at $|z| >
2H$ we find features in the magnetic field correlation function on
scales of $\sim 10H$, indicating the existence of meso-scale structure.
Although the magnitude of the mass, angular momentum and energy
transport in the corona is small compared to that in the central disk,
the corona and the central disk are dynamically connected and large
scale structure in magnetically dominated upper layers may still
influence the spatial correlations/structures of the disk below (we will
explore this issue in a forthcoming paper). Therefore, in accretion disk
models where the spatial structure of the corona is important, such as
phenomenological models for accretion disk coronae (e.g., the
statistical model of \citealt{uzgoo}) and disk spectra calculations, the
radial extent of the corresponding numerical simulation may require a
$L_x \geq 10 H$. 

(3) $L_z$. We have investigated the effect of vertical boundaries by
running a model (s16c) with $L_z = 12H$. We find no qualitative
difference between the $L_z = 12H$ and $L_z = 10H$ models: the
saturation $\overline{\<\alpha\>} \sim 0.0141$, $\overline{\<E_{B, {\rm
d}}\>}/\rho_{0}c_s^2 \sim 0.0125$, and  $\overline{\<E_{B, {\rm
c}}\>}/\rho _{0}c_s^2 \sim 0.00497$, all within $\sim 10\%$ of the
fiducial model. We also obtain a similar vertical disk structure when
$L_z$ increases: inside $2.5H$ the disk has a well-fitted Gaussian
$\overline{[\rho]} (z)$ and a flat $\overline{[E_B]}(z)$; outside $2.5H$
the corona extends to  $|a| = \pm 6H$ in model s16c. Both
$\overline{[\rho]} (z)$ and $\overline{[E_B]}(z)$ also have an
exponential profile, but the fitted coronal exponential scale height
increases $\sim 20\%$ compared to the fiducial model. The coronal loop
distribution functions are almost identical to those of the fiducial
model, which is not suprising considering the steep decline of
${d\Phi}/{d \Delta z_{\rm max}}$ as $\Delta z_{\rm max}$ exceeds $\sim
H$.  As discussed before, some caution is needed in interpreting this
vertical extension of corona structure with increasing domain size: our
calculation is essentially a MHD calculation, whereas real disk coronae
are probably force-free and also influenced by non-ideal plasma effects
(e.g., reconnection) that are ill-modeled in our numerical scheme.

(4) $\beta _{0}$. We have tested the effect of initial field strength on
the saturation level. In most of our runs we start from a uniform
toroidal field inside the disk with $\beta _ 0 = 25$. We then carry out
a comparison run s16b with the same field geometry but weaker strength
$\beta _ 0 = 100$. We find that turbulence saturates at the similar
level using weaker initial field strength, with $\overline{\<\alpha\>}
\sim 0.0157$, $\overline{\<E_{B, {\rm d}}\>}/\rho_{0}c_s^2 \sim 0.0152$,
and $\overline{\<E_{B, {\rm c}}\>}/\rho _{0}c_s^2 \sim 0.00647$,
therefore in stratified disks the saturation does not depend on the
initial field strength. 

In comparison, in unstratified disk models $\<E_B\>$ is found to scale
with the initial mean field strength $\<B_y\>$\footnote{\cite{ggsj09}
found a linear relation between $\<B_y\>_d$ and the saturation
$\overline{\alpha} \propto \overline{\<E_B\>} \propto \rho_0 c_s
V_{A,y0}$, where $V_{A,y0} = B_{y0}/(4\pi \rho_0 c_s^2)^{1/2}$ is the
mean azimuthal Alfv\'en speed.  This result is also consistent with
scalings obtained from earlier work (e.g., \cite{hgb95}).  } (see a
detail discussion in \citealt{ggsj09}).  The important difference here
lies in the stratification and the accompanying outflow boundary
conditions, which allow changes in mean toroidal field strength in the
turbulent disk.  The stratified disk model then allows the disk to
adjust its net flux and field strength in a self-consistent way. It is
worth pointing out that the saturation level is a volume average over
large scales where different parts of the disk decorrelate; on the scale
where the turbulence is localized ($\leq H$), it is still possible that
the local saturation $\<E_{B,{\rm d}}\>_{\rm local} \propto \<B_{y, {\rm
d}}\> _{\rm local}$.

\begin{figure}
\plotone{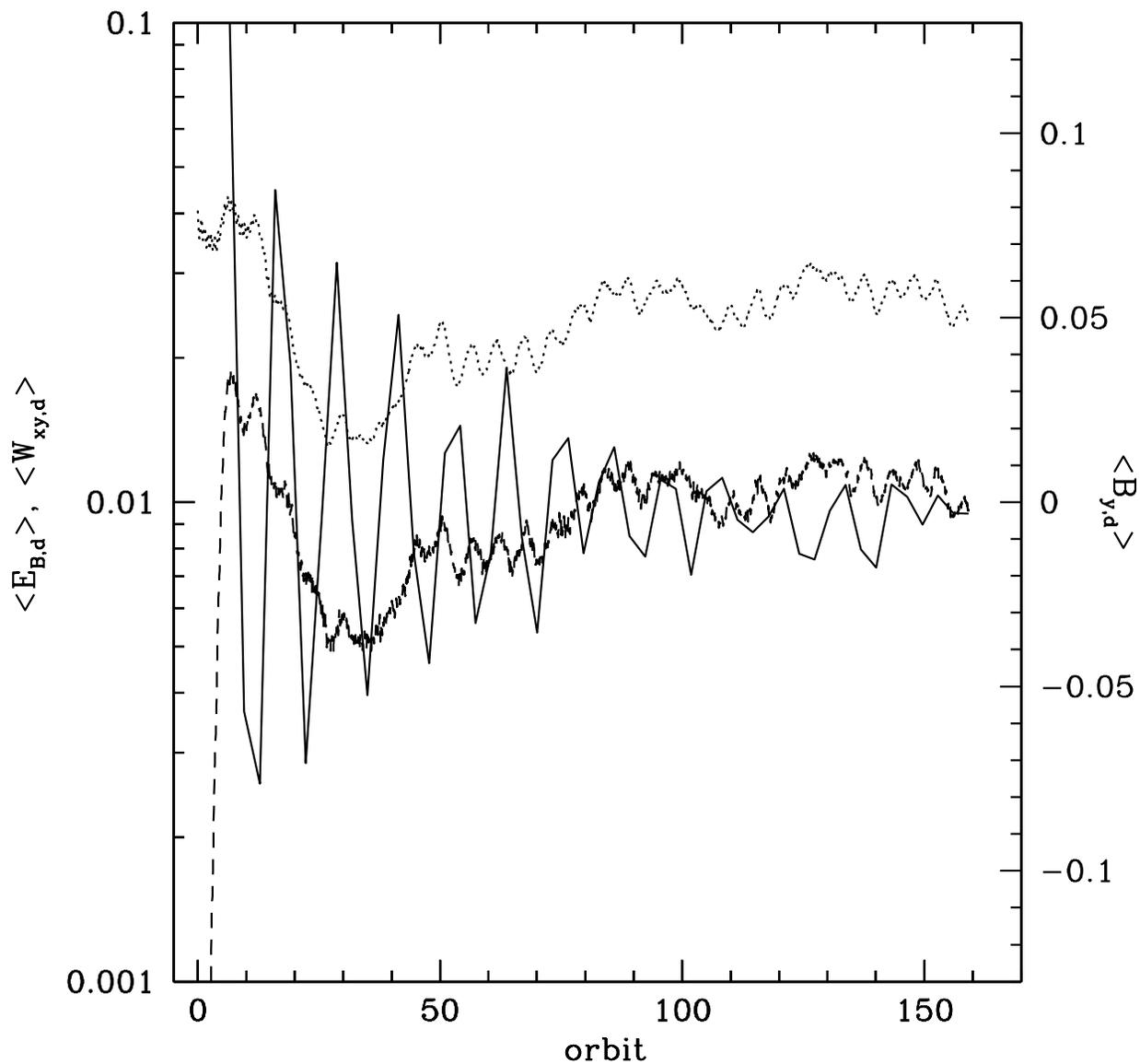}
\caption{ Evolution of $\<B_y\>$ (solid lines), $\<E_B\>$ (dotted
  lines) and $\<W_{xy}\>$ (dashed lines) at $|z| \leq 2H$ in model s32. The
oscillation period for $\<B_y\>$ is twice that of $\<E_B\>$ and
$\<W_{xy}\>$. These temporal oscillations may be caused by a mean field dynamo.
}
\label{fig:s32.mb.vs.z}
\end{figure}

Does the saturation level depend on the instantaneous mean field
strength in the stratified disk? The evolution of the mean azimuthal
field $\<B_{y,{\rm d}}\>$ in the region $|z| \leq 2H$ in model s32 is
plotted in Figure \ref{fig:s32.mb.vs.z}. The mean field does not have a
fixed value and changes signs over a time scale of $\sim 10$ orbits.
Averaging the last $100$ orbits, the mean magnitude of toroidal field
strength in the turbulent disk region is $\overline{|\<B_y\>|} \sim
0.012\sqrt{4\pi} \rho_0^{1/2}c_s$, therefore there is weak net toroidal
field in the turbulent disk region to drive MRI. The figure also shows
the evolution of the mean magnetic energy density $\<E_{B,{\rm d}}\>$
and mean $xy$ stress $\<W_{xy,{\rm d}}\>$ in the same run. Both
$\<E_{B,{\rm d}}\>$ and $\<W_{xy,{\rm d}}\>$ have a fixed overall
saturation level with a superimposed small oscillation with a period
half of that of $\<B_y\>$.  Again, this is dramatically different from
the unstratified disk, where the saturation level is proportional to the
mean field strength. The overall saturation level in a stratified disk
is ${\it not}$ determined by the instantaneous mean field strength, nor
by the initial field strength. 

On the other hand, as shown in Figure \ref{fig:s32.mb.vs.z}, the
oscillations of $\<B_y\>$ and $\<E_{B,{\rm d}}\>$ are closely correlated
and oscillation period for $\<B_y\>$ is twice that of $\<E_{B,{\rm
d}}\>$. This suggests that $\<B^{2}\>$ is correlated with $\<B_y\>$,
even though $\<B_y\> \ll \<B^2\>^{1/2}$. Therefore, the saturation level
may be determined both by the MRI induced by the mean toroidal field in
the turbulent disk region and magnetic buoyancy effects (e.g.,
\cite{v09}). 

\section{Butterfly Diagram: A Mean Dynamo In the Disk?}

\begin{figure}
\plotone{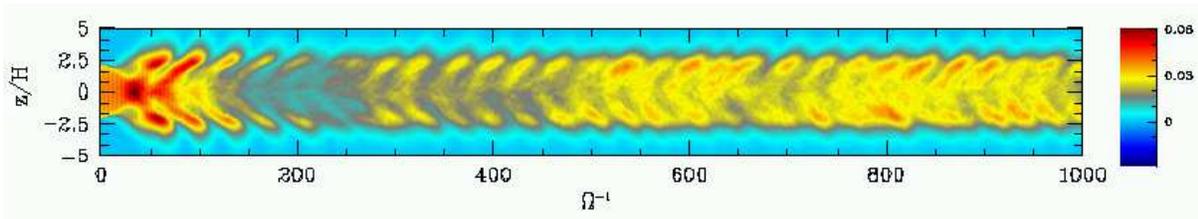}
\caption{Butterfly diagram for $[E_B]$ in the model s32. }
\label{fig:butterfly}
\end{figure}

One interesting feature appearing in all our models is an oscillation of
the mean magnetic energy on a timescale of a few orbits. As an example,
we plot the ``butterfly" diagram for model s32 in Figure
\ref{fig:butterfly}, which illustrates the evolution of $[E_B](z)$. This
bears a superficial resemblance to the famous butterfly diagram observed
in solar activity cycles. 

\begin{figure}
\plotone{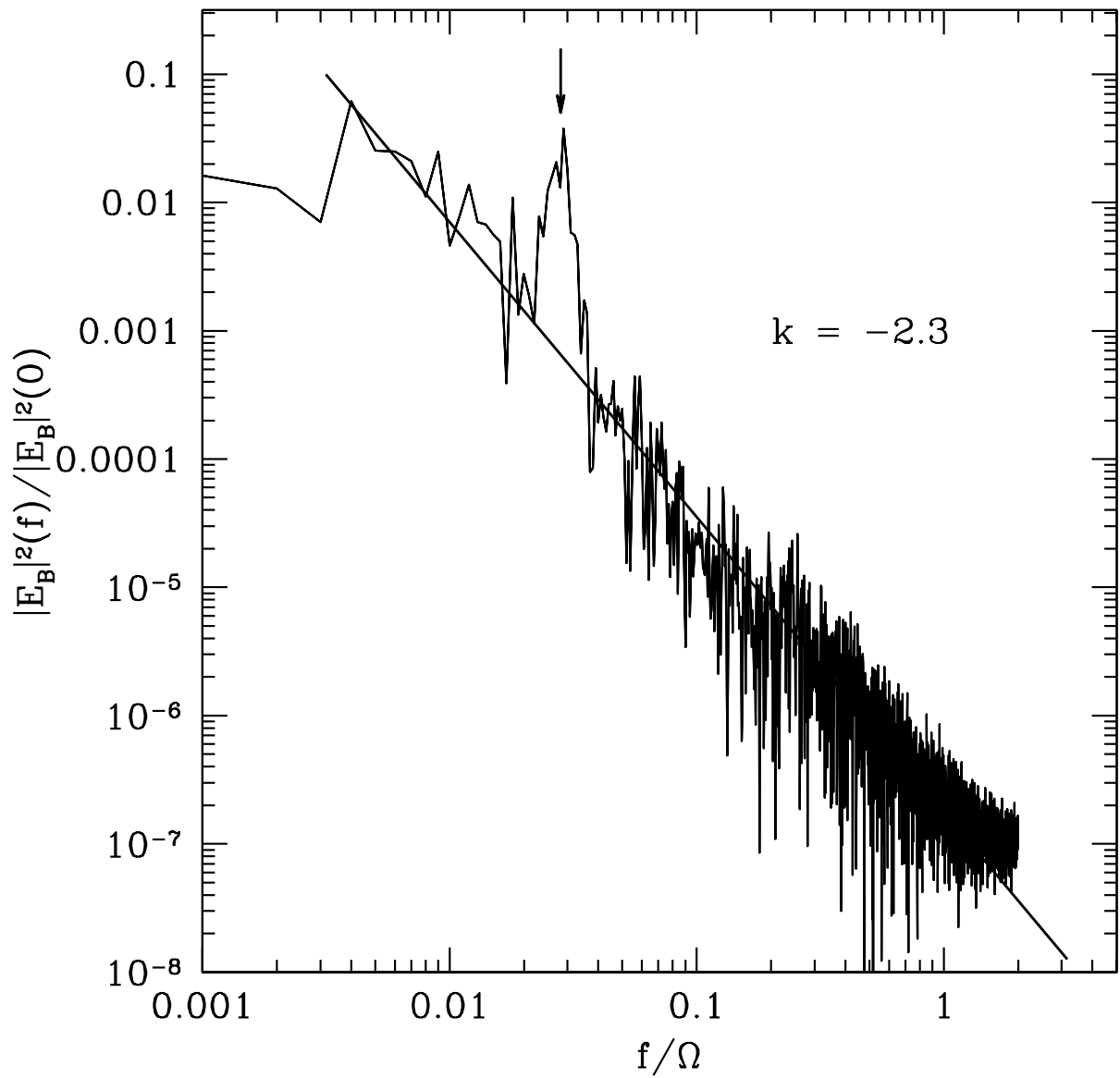}
\caption{Normalized temporal power spectral density for $[E_B]$ in the model
  s32. The data are taken from the layers with $|z| \sim 2.5 H$. We
  also draw the best-fit $ k = -2.3$ slope for the temporal PSD. The
  arrow marks the peak frequency in the power spectrum. This
  frequency,  $f \sim 0.03 \Omega$, corresponds to the period of the butterfly
  diagram for $[E_B]$. }.
\label{fig:butterfly.pow}
\end{figure}

We use Fourier analysis to determine the period of butterfly diagram.
Using data $\int dy E_B(|z| = 2.5H; x,t)$, taken from two layers with
$|z| \sim 2.5H$ and have been averaged in $y$ direction to improve
statistics, we perform a two dimensional FFT (in $x$ and $t$) on the
data set. The normalized temporal power spectral density (PSD) for
$[E_{B, |z|=2.5H}]$ in model s32 are shown in Figure
\ref{fig:butterfly.pow}. Here we have plotted a cut through $k_x = 0$
plane in the 2D $k_x-f$ PSD map. We have also checked that the different
sides of the disk have very similar PSD and we have plotted the sum of
contribution from both layers. The arrow in the figure marks the peak
frequency in the PSD. This frequency, $f \sim 0.03\Omega$ corresponds to
the period of the butterfly diagram for $[E_B]$. 

The PSD has $P\sim f^{k}$, with $k \sim -2.3$. Interestingly, results
from recent global GRMHD simulations \citep{nk09} suggest that the slope
for the coronal luminosity temporal power spectrum is $k \sim -2$,
almost independent of model parameters and very close to what has been
observed at high frequency in black hole accretion disk systems. The
power-law index for the temporal power spectrum from local and global
simulations are therefore remarkably close, considering we are only
calculating the temporal spectrum for coronal magnetic energy density.

The period for $[E_B]$ is $P_{[E_B]} \sim 5$ orbits. Besides $[E_B]$,
one could also plot butterfly diagrams for $[W_{xy}]$ and $[B_y]$. The
period for $[W_{xy}]$ is the same as that of $[E_B]$, $\sim 5$ orbits,
while the period for $[B_y]$ is twice that of $[E_B]$, $ P_{[B_y]} \sim
10$ orbits, because of the reversal of mean fields (see Figure
\ref{fig:s32.mb.vs.z}). 

For $[E_B]$, we find $P_{\<E_B\>}\sim 5$ orbits in all our models. This
quasi periodicity has appeared in all the stratified shearing box
simulations that we are aware of, even in those with periodic vertical
boundary conditions (Stone et al., 2009). Interestingly, \cite{rf08} has
also obtained similar butterfly diagrams at certain radii (e.g., $r = 8
r_g$ and $r = 10 r_g$, where $r_g \equiv GM/c^2$ is the gravitational
radii) in their global pseudo-Newtonian thin disk simulations. It would
be of interest in the future to test (a) whether the butterfly diagram
is simply a local feature at a certain location on the disk (as in
shearing box simulations) or this quasi-periodicity can be coherent and
sustained over a large radial range, and (b) what model parameter(s) the
period depends on.  

The butterfly diagram together with the reversal of the mean fields (for
both the dominant toroidal field and a weak radial field) in the disk
may be modeled by a mean field dynamo of $\tilde{\alpha}$ type (e.g.,
\cite{moffatt78}\footnote{In this work we use $\tilde{\alpha}$ to denote
dynamo model type. It should not be confused with the accretion disk
turbulence level parameter $\alpha$ }). In the rest of this section, we
will present a toy model to give a qualitative description of these
oscillations.

Let us first consider two important dynamical processes in a stratified
disk: (1) the MRI-driven turbulence, which draws free energy of rotation
and operates on the orbital timescale $\sim \Omega ^{-1}$; (2) magnetic
buoyancy, which operates on the local Alfv\'en timescale $\tau_{\rm A}
\sim H/[\delta v_{A}^2]^{1/2}$. In our simulations we found in the disk
region $|z| \leq 2H$ the magnetic energy density is almost a constant
with height, with $[E_B] \geq 10^{-2} \rho _0 c_s^2 $, which gives an
average magnetic buoyancy timescale $\tau_{\rm A} \leq$ a few $\Omega
^{-1}$ inside the disk. The period of the butterfly diagram is much
longer than these two timescales. Therefore these two processes alone
can not describe the dynamics represented in butterfly diagrams.

The $\tilde{\alpha}$ type mean dynamo equations for the disk mean fields
$\<B_x\>$ and $\<B_y\>$ can be derived from averaging the induction
equation, $\partial_t{\<\bB\>} = \<\nabla \times ({\bv}\times{\bB})\>$,
here $\<\,\,\>$ denotes ensemble averages. Assuming the turbulent EMF
$\bld{\varepsilon}$ is related to the mean field with a dynamo parameter
$\tilde{\alpha}_{i}$, $\<{\bld{\varepsilon}}\> \equiv \<{\delta
\bv}\times {\delta \bB}\> =  \tilde{\alpha}_{i} \<{\bB}_i\>$, one simple
form of dynamo equations in a stratified thin Keplerian disk is (cf.,
Eqn (5-6) in \cite{vb97}),
\begin{equation}
\del_t \<B_y\>   = -\frac{3}{2}\Omega \<B_x\> - \del_z(\<v_{\rm b}\>\<B_y\>) + \del_z(
\tilde{\alpha}_{1}\<B_x\>)\,\,,
\label{eqn:dynamo.by0}
\end{equation}
and 
\begin{equation}
\del_t {B_x} =  -\del_z(\<v_{\rm b}\>\<B_x\>) - \del_z(\tilde{\alpha}_{2}\<B_y\>)\,\, ,
\label{eqn:dynamo.bx0}
\end{equation}
where $v_{\rm b}$ is a characteristic vertical velocity induced by
magnetic buoyancy. In Eqn (\ref{eqn:dynamo.by0}) the first term is the
shear term, second term denotes buoyancy due to the mean field, and the
last term is the mean field dynamo term. Only $\del_z$ terms are
retained because the disk is thin. For simplicity we have also dropped
the diffusion terms.  We then take $\partial_z \sim 1/(2H)$ and $v_{\rm
b} \approx |v_A| \equiv |B|/\sqrt{4\pi\rho_0} \sim
|B_y|/\sqrt{4\pi\rho_0}$, where $|v_A|$ is the mean Alfv\'en speed.
Eqn(\ref{eqn:dynamo.by0}) and Eqn(\ref{eqn:dynamo.bx0}) then become
\begin{equation}
\frac{dB_y}{dt} = -\frac{3}{2}\Omega B_x - \frac{|v_A|}{2H}B_y + \frac{
\tilde{\alpha}_{1}}{2H}B_x\,\,,
\label{eqn:dynamo.by}
\end{equation}
and
\begin{equation}
\frac{dB_x}{dt} =  -\frac{|v_A|}{2H}B_x - \frac{
\tilde{\alpha}_{2}}{2H}B_y\,\,.
\label{eqn:dynamo.bx}
\end{equation}
For clarity we have dropped $\<\,\,\>$ in the above equations. Notice 
that Eqn (\ref{eqn:dynamo.by}) and Eqn (\ref{eqn:dynamo.bx}) have no spatial
dependence. Taken together, they are coupled ODEs
and can be solved numerically given initial conditions for
$B_y$ and $B_x$. 

\begin{figure}
\plotone{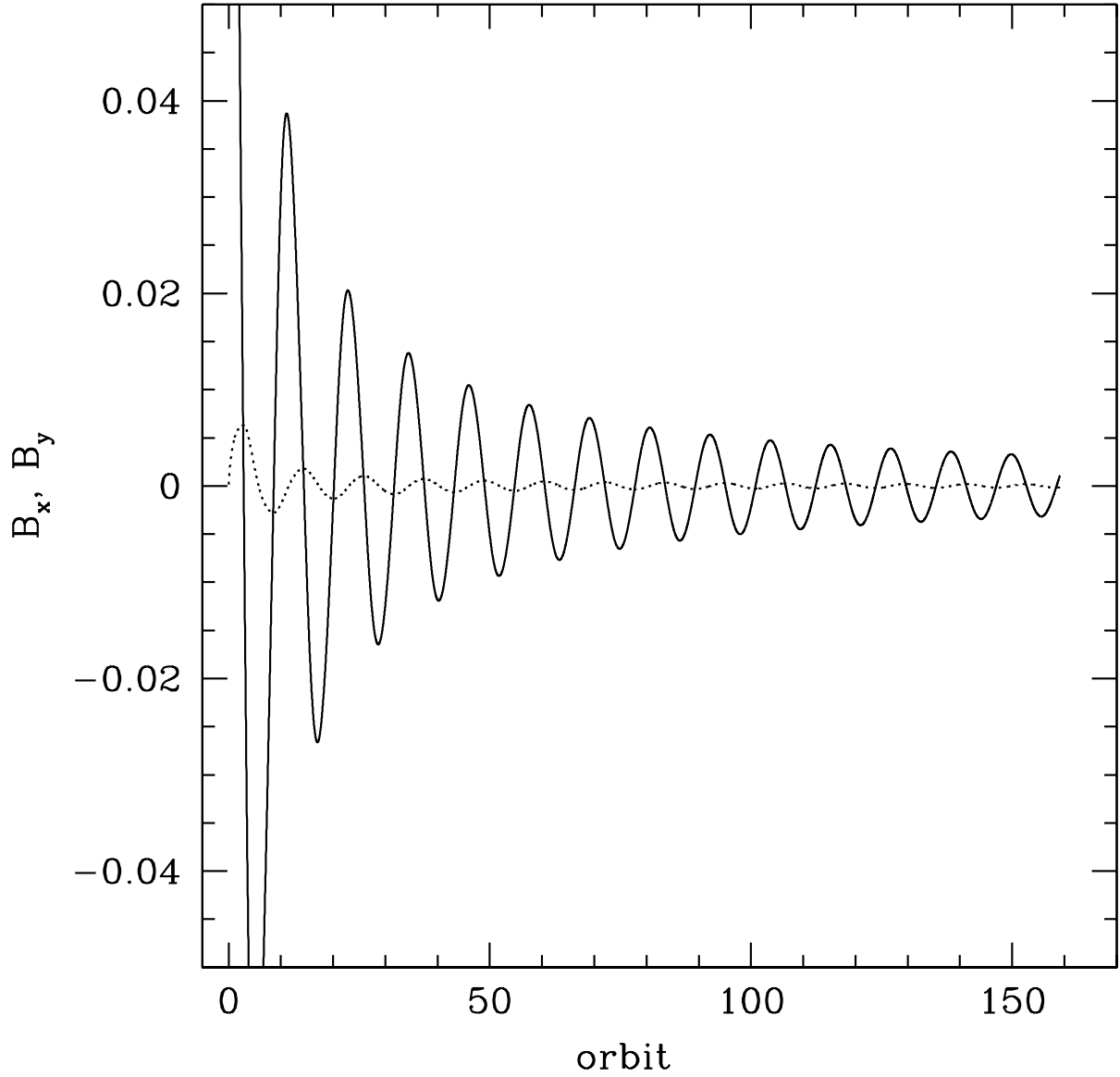}
\caption{ Evolution of $B_y$ (solid lines) and $B_x$ (dotted lines) in our mean field dynamo toy model. $\alpha _{1} = \alpha _{2} = -0.01$ in the plotted model. }.
\label{fig:toymodel}
\end{figure}

In Figure \ref{fig:toymodel} we plot
one solution for this toy model. This solution is obtained by
integrating the above equations from an initially pure
toroidal field with $\beta_{0} \sim 22$ and by choosing $\tilde{\alpha} _{1} = \tilde{\alpha} _{2} = -0.01$\footnote{By definition, $\tilde{\alpha}_{1} =
  \frac{\<\varepsilon _x\>}{\<B_x\>} =\frac{\<\delta v_y\delta B_z -
    \delta v_z\delta B_y\>}{\<B_x\>}$, $\tilde{\alpha}_{2} =
  \frac{\<\varepsilon _y\>}{\<B_y\>} = \frac{\<\delta v_z\delta B_x -
    \delta v_x\delta B_z\>}{\<B_y\>}$. In principle, $\tilde{\alpha}_1$ does
  not necessarily equal $\tilde{\alpha}_2$ due to anisotropy. 
}.  The period for $B_y$ in this particular toy model is $\sim 10$ orbits.
The magnitude of $\tilde{\alpha}$ controls the oscillation
frequency: in general, larger $|\tilde{\alpha}|$ leads to smaller
period, although the scaling is not linear. Initial conditions have
little effect on the evolution in our toy model. In conclusion, the
butterfly diagram and the mean field reversal observed in these
simulations may imply a mean field dynamo at work in stratified disks.

Does it make sense to identify these oscillations with observed QPOs?
In a Keplerian disk the orbital frequency at $r$ is $f_{\rm orb} =
1/(2\pi)(GM)^{1/2}r^{-3/2}$.  The QPO frequency is $f_{\rm QPO} = 1/5
f_{\rm orb} \approx 20 \times (r/10M)^{-3/2}(M/10\msun)^{-1}{\rm Hz}$.
Our disk model represents a geometrically thin, optically thick disk.
This is most easily understood as corresponding to the high soft state
in black hole X-ray binaries, which is dominated by a thermal component.
For a $10\msun$ black hole a $5{\rm Hz}$ QPO (e.g., XTEJ1550-564)
corresponds to $r_{\rm in} \sim 25M$, which is far from innermost region
of a thin disk where most of the thermal X-ray emissions presumably
originates.  This oscillation frequency may be sensitive to the disk
vertical structure (e.g. if the disk is not isothermal), and therefore
may exhibit a much more complex behavior in real disks, in which the
vertical structure is closely coupled to vertical energy transport. On
the other hand, observations indicates that QPOs are absent or very weak
in the thermal state, but may appear in the very high state when a
sizable thermal disk component is present, although the QPOs are more
associated with Comptonizing electrons (\citep{rm06}); it is difficult
to associate the butterfly oscillations with observed QPO phenomena. 

\section{Summary and Discussion}

We have carried out stratified shearing box simulations with domain size
$L_x = H$ to $L_x = 32H$ to study properties of isothermal accretion
disks on a scale larger than the disk scale height $H$.  Our numerical
models have vertical extent $\geq 5H$ above and below the disk midplane
with outflow boundary conditions. All models start from a net mean
toroidal field in the central disk region and the mean fields are
allowed to change in the evolution.  

We find the disk has an oscillating mean toroidal field and
$\overline{\<\alpha\>} \sim 0.012 - 0.025$ in the parameter range we
explored. We have not found a clear dependence of
$\overline{\<\alpha\>}$ on $L_x$ in our models, although the temporal
variances in volume averaged quantities decreases with $L_x$. The
highest resolution used here is modest ($20-40$ zones per $H$), and we
have observed $\overline{\<\alpha\>}$ increases with resolution.
Recently, Stone el al. report a converged $\overline{\<\alpha\>} \sim
0.04$ in $L_x \sim H$ high resolution stratified disk simulations with
zero-net-flux and periodic vertical boundary conditions (so that the
volume-averaged field cannot change during the evolution).  The
sustained turbulence may be due to the presence of a mean toroidal
field in the region close to the disk midplane, lending plausibility to
the idea that the saturation mechanism of MRI in stratified disks near
the midplane is similar to that in unstratified disks with a net
toroidal field.

In the saturated state the disk vertical structure consists of (a) a
turbulent disk at $|z| \leq 2H$ and (b) a magnetically dominated upper
region at $|z| > 2H$, confirming earlier small ($L_x \sim H$) box
results.  

At $|z| \leq 2H$, the disk is mainly supported by gas pressure, and a
Gaussian density profile is observed. The plane averaged magnetic energy
density $[E_B](z)$ and Maxwell stress $[M_{xy}](z)$ are nearly uniform
with vertical height $z$ in this region, where the disk is marginally
stable to the Parker instability. At $|z| > 2H$, exponential dependences
on $z$ are observed for both $[\rho]$ and $[E_B]$. Fitting formulae for
$[\rho](z)$ and $[E_B](z)$ are given in Eqn (\ref{eqn:rhoz}) and Eqn
(\ref{eqn:ebz}) respectively. 

Using a two-point correlation function analysis, we found that close to
the midplane, the disk is dominated by small scale ($\leq H$)
turbulence, very similar to what we have observed in unstratified disk
models. In the corona, magnetic fields are correlated on scales of $\sim
10H$, implying the existence of meso-scale structures. Recently
\cite{jky09} have also observed large scale pressure and zonal flow
structures in their large shearing box simulations. We will give a
detailed report of meso-scale structure in isothermal disks in a
forthcoming paper.

We have adopted a statistical approach to study the geometry of coronal
magnetic fields.  Only $\approx 4\%$ of coronal field lines are open.
For closed field lines, we calculated the magnetic loop distribution
function for the loop foot separation $\Delta \bx$ in the $x-y$ plane,
loop maximum height $\Delta z_{\rm max}$, and loop orientation angle
$\theta _{\rm foot}$.  The loops are dominantly toroidal due to the
differential shear. The loop foot distribution between $H-20H$ is a
power law with an index $k \sim -1.25$. In the phenomenological model of
UG, this corresponds to the limit where reconnection is slow compared to
the shear. These comparisons are limited because our models are working
in an ideal MHD regime and reconnection is purely numerical.

In our models both vertical energy and momentum flux are negligible in
the steady state. The mass loss rate from the disk surface is small and
decreases with increasing $L_z$. The surface effects are therefore
minimal and indicate a lack of disk winds in our stratified disk models.
The weak winds are consistent with the constraint that we have a
zero-net vertical magnetic flux in these models. A Blandford-Payne type
wind requires the existence of a vertical net field (e.g., see
\cite{si09}), although we note that in their models the most unstable
wavelength for the extremely weak field  are probably not resolved.)
Initial investigations show that even a weak ($\beta _{0} \sim 1600$)
net z field will induce very violent accretion in stratified shearing
box models: at certain region of the disk accretion will run away,
eventually causing the disk break into rings.  Similar phenomena were
reported in net vertical field models of \cite{ms00}.

We have confirmed the ``butterfly" diagram seen in earlier
stratified disk models of size $L_x \sim H$. The butterfly diagrams 
persist even in our largest runs with $L_x = 32H$.  We
also report the reversal of the mean fields (for both the dominant
toroidal field and a weak radial field) in the disk on a timescale twice
that of $[E_B]$. The periods for the butterfly diagram are close in all
our models, $P \sim 5$ orbits for $[E_B]$ and $P \sim 10$ orbits for
$[B_y]$. The mean field reversal and butterfly diagram may indicate the
existence of a mean field dynamo in stratified disks, perhaps controlled
by the MRI and magnetic buoyancy. We have presented a toy model for an
$\tilde{\alpha}$ type mean field dynamo in stratified disks
and found an $\tilde{\alpha} _{\rm imp} \sim 0.01$ will produce the reported
period. Further exploration of parameter dependences would be useful for
analytical modeling.  In the future it would also be interesting to test
whether the butterfly oscillations persist when averaging over a
large range of radii in global disk simulations. The butterfly
diagram may be associated with low frequency QPOs and therefore a
good observational diagnostic for accretion flows. On the other hand, we
also report a power-law index $k \sim -2.3$ in the temporal power
spectrum for coronal magnetic energy fluctuations, consistent with
results from recent GRMHD black hole accretion disk simulations.

Our stratified disk models are primarily limited by the assumption that
the disk is isothermal. Effects of thermodynamics and radiation
therefore are neglected in this work. Our models are also limited by
finite resolution, box size, evolution time, and the absense of explicit
dissipation. Additional insights may also provided by the future
explorations on magnetic field strength and geometry in disks.

\acknowledgements

This work was supported by the National Science Foundation under grants
AST 00-93091, PHY 02-05155 and AST 07-09246, and a Sony Faculty
Fellowship, a University Scholar appointment, and a Richard and Margaret
Romano Professorial Scholarship to CFG, and by the NASA under grant
NNX09AD14G to John Hawley. The authors are grateful to John Hawley, Stu
Shapiro, Bill Watson, and Jake Simon for discussions. Simulations were
performed at following NSF supported Teragrid cites: \abe/NCSA,
\queenbee/LONI, \ranger/TACC, and \kraken/NICS.

\newpage
\begin{deluxetable}{lcccccc}
\setlength{\tabcolsep}{0.07in}
\tabletypesize{\scriptsize}
\tablecolumns{7}
\tablecaption{Model Parameters
  \label{size}}
\tablewidth{0pt}
\tablehead{
\colhead{Model} & \colhead{Size} & \colhead{Resolution} & \colhead{$\beta_0$} &
\colhead{$\overline{\<\alpha\>}$} & \colhead{$\overline{\< E_{B, |z|\leq
      2H} \>}/\rho_{0}c_s^2$ } & \colhead{$\overline{\< E_{B, |z| > 2H} \>}/\rho_{0}c_s^2$}  \\
}
\startdata
std16 & $(16, 20, 10)H$ & $384\times 256\times 128$ & 25 & 0.0125 &
0.0121 & 0.00427 \\
s16b  & $(16, 20, 10)H$ & $384\times 256\times 128$ & 100 & 0.0157 &
0.0152 & 0.00647 \\
s16c  & $(16, 20, 12)H$ & $384\times 256\times 160$ & 25 & 0.0141 &
0.0125 & 0.00497 \\
s1 & $(1, 20, 10)H$ & $48\times 256 \times 128$ & 25 & 0.0191 & 0.0171
& 0.00933 \\
s8 & $(8, 20, 10)H$ & $192\times 256 \times 128 $ & 25 &
0.0124 & 0.0115 & 0.00665 \\
s32 & $(32, 20, 10)H$ & $768\times 256 \times 128$ & 25 &
0.0269 & 0.0270& 0.0106 \\
s16a  &  $(16, 20, 10)H$ & $768\times 512\times 256$ & 25 & 0.0230 &
0.0181 & 0.0101 \\

\enddata
\end{deluxetable}

\clearpage

\clearpage

\clearpage

\clearpage

\clearpage

\clearpage

\clearpage

\clearpage

\clearpage

\clearpage

\clearpage

\end{document}